\documentclass[showpacs,showkeys,preprintnumbers,amsmath,amssymb,twocolumn,pra,floatfix]
 {revtex4}

\usepackage{amssymb,amsfonts}
\usepackage{amsmath}
\usepackage{hyperref}
\usepackage{graphicx}
\usepackage[usenames,dvipsnames]{xcolor}
\graphicspath{{Figures/}}
\begin{document}

\title{Quench dynamics and defects formation in the Ising chain in a transverse magnetic field}

\author{Alexander I Nesterov}%
  \email{nesterov@cencar.udg.mx}
\affiliation{Departamento de F{\'\i}sica, CUCEI, Universidad de Guadalajara,
Av. Revoluci\'on 1500, Guadalajara, CP 44420, Jalisco, M\'exico}

\author{M\'onica F Ram\'irez}
 \email{monica.felipa@gmail.com}
 \affiliation{ Tepatit\'an’s Institute for Theoretical Studies,
 Tepatit\'an, Jalisco, M\'exico}

\date{\today}

\begin{abstract}
We study analytically and numerically  quench dynamics and defects formation in 
the quantum Ising model in the presence of a time-dependent transverse magnetic field. We generalize the Landau-Ziner formula to the case of non-adiabatic evolution of the quantum system. For a quasi-static magnetic field, with a slow dependence on time, our outcomes are similar to the results predicted by the Landau-Zener formula. However, a  quench dynamics under a shock-wave load is more complicated. The final state of the system  depends on the amplitude and pulse velocity, resulting in the mixture of ground and excited states and significant density of defects.
\end{abstract}


\maketitle

\section{Introduction}

Quantum phase transitions (QPTs) are characterized by qualitative changes 
of the ground state of many body system and occur at the zero temperature 
\cite{SS}. Since thermal fluctuations are frozen, QPTs are purely quantum 
phenomena driven by quantum fluctuations. Well known examples of QPTs 
are the superconductor to insulator transition in high-$T_c$ 
superconductoing systems, the quantum paramagnet to ferromagnet 
transition occuring in Ising spin system under an external transverse 
magnetic field, and the superfluid to Mott insulator transition.

QPTs are associated with levels crossing and imply the lost analyticity in the 
energy spectrum. In the parameter space the points of non-analyticity, being 
referred to as critical points, define the QPT \cite{SS}. A first-order QPT is 
determined by a discontinuity in the first derivative of the ground state 
energy. A second-order QPT means that the first derivative is continuous, 
while the second derivative has either a finite discontinuity or divergence at 
the critical point.

Energy level crossing implies that it does not matter how slowly the system 
is evolved. Near the critical point adiabaticity breaks down and 
non-equilibrium phenomena associated  with the drastically grown quantum 
fluctuations can drive the system away from the ground state. The final 
result depends on how fast the transition  occurs. If the quench process is 
sufficiently fast, large numbers of topological defects are created and the 
final state can be essentially different from that been obtained as result of 
slow evolution. Qualitatively, the  dynamics of quantum system can be 
described by the  Kibble-Zurek (KZ) theory of nonequilibrium phase 
transitions \cite{KTW,ZHW,ZHW1}.

In this paper we consider quench dynamics of the quantum of Ising chain in 
a transverse time-dependent magnetic field. The paper is organized as follows: In Sec. II 
we introduce the model and discuss its main features. In Sec. III we study 
quench dynamics for a magnetic field defined by a pulse of a given shape. In 
Sec. IV  we describe defects formation near of critical point. We conclude in 
Sec. V with a discussion of our results.

\section{Description of the model}

We consider the one-dimensional Ising model in a transverse magnetic field governed by the following  Hamiltonian:
\begin{align} \label{EqH1}
{\mathcal H}= - \frac{J}{2}\sum^N_{n=1}\big(h \sigma^x_n + \sigma^z_n \sigma^z_{n+1} ),
\end{align}
where  the periodic boundary conditions, $\mbox{\boldmath$\sigma$}_{N+1} =\mbox{\boldmath$ \sigma$}_1$, are imposed. The external magnetic field is associated with the parameter $h$. Quantum phase transition (QPT) occurs in the thermodynamic limit ($N \gg 1$) at critical value $h_c =1$ of the external magnetic field.

The Hamiltonian in Eq. (\ref{EqH1}) can be diagonalized using the standard Jordan-Wigner transformation, following well-known procedures described in \cite{LSM,KSH,SMAL,MBM1,BMBM}.
The Jordan-Wigner transformation maps a spin-1/2 system to a system of spinless fermions, 
\begin{align}\label{Eq1c}
\sigma^x_n = &1-2 c^\dagger_n c_n, \\
\sigma^y_n = &i(c^\dagger_n - c_n  )\prod_{m<n}(1-2 c^\dagger_m c_m), \\
\sigma^z_n =& -(c_n + c^\dagger_n )\prod_{m<n}(1-2 c^\dagger_m c_m),
\end{align}
with anticommutation relations:
 $\{c^\dagger_m,c_n\} = \delta_{mn}$ and $\{c_m,c_n\}=\{c^\dagger_m,c^\dagger_n\}=0$.
Applying these transformations, we obtain
\begin{widetext}
\begin{align}\label{I}
{\mathcal H} = -\displaystyle\frac{J}{2}\sum^N_{n=1}\big(c^\dagger_n c_{n+1 }+ 
c^\dagger_{n+1}c_{n}+ c^\dagger_{n} c^\dagger_{n+1} 
+ c_{n +1}c_{n }+ h(1-2 c^\dagger_n c_{n })\big).
\end{align}
\end{widetext}
The periodic boundary conditions imposed on the spin operators lead to the following condition for the fermionic operators:
 \begin{align}\label{Eq1b}
  c_{N+1} = - e^{i\pi {\mathcal N}_F} c_1,
 \end{align}
${\mathcal N}_F = \sum^N_{n=1}  c^\dagger_{n} c_{n}$ being the total number of fermions. Using Eq. (\ref{Eq1c}) we find that ${\mathcal N}_F=  N/2- S^x$, where $S^x= (1/2)\sum_n  \sigma^x_n$ is the total $x$-component of the spins. For the particular choice of $S^x =0$,  we obtain ${\mathcal N}_F=  N/2$. This yields periodic {\em periodic} ({\em antiperiodic}) boundary conditions for $c_{n} $,  if $N/2$ is {\em odd} ({\em even}). Since the parity of the fermions is conserved, the imposed boundary conditions are valid for all values of the parameter $h$.

Applying the Fourier transformations,
\begin{align}
c_n = \frac{e^{-i\pi/4}}{\sqrt{N}}\sum_k c_k e^{i2\pi kn/N},
\end{align}
we find that the Hamiltonian (\ref{I}) can be recast in Fourier space as
\begin{widetext}
\begin{align}\label{Ham}
{\mathcal H}= \frac{J}{2}\sum_{k}\Big (2(h-\cos\varphi_k)c^\dagger_k c_k-  h 
+ \sin \varphi_k(c^\dagger_k c^\dagger_{-k} + c_{-k} c_{k} ) \Big),
\end{align}
\end{widetext}
where $\varphi_k = {2\pi k}/{N}$. For periodic boundary conditions, $c_{N+1}=c_1$, the wave number $k$ takes the following discrete values:
  \begin{align}\label{Eq1ar}
 k=-\frac{N}{2}, \dots, 0,1, \dots,  \frac{N}{2}-1,
 \end{align}
and for antiperiodic boundary conditions, $c_{N+1}=-c_1$, one has  \begin{align}\label{Eq1a}
 k=\pm\frac{1}{2},\pm\frac{3}{2},\dots, \pm\frac{N-1}{2}.
  \end{align}
 Here we set lattice spacing $a=1$. In what follows, we impose the
antiperiodic boundary conditions for the fermionic operators.

The Hamiltonian (\ref{Ham}) can be diagonalized by using the Bogoliubov transformation,
\begin{align}\label{Eq6d}
&c_k = \cos\frac{\theta_k}{2} \, a_k +  \sin\frac{\theta_{-k}}{2} \,a^\dagger_{-k},  \\
& c^\dagger_k =  \cos\frac{\theta_k}{2} \,a^\dagger_k +  \sin\frac{\theta_{-k}}{2} \,a_{-k}, \label{Eq6b}\\
&a_k = \cos\frac{\theta_k}{2} \, c_k +  \sin\frac{\theta_k}{2} \,c^\dagger_{-k}, \label{Eq6c}\\
& a^\dagger_k = \cos\frac{\theta_k}{2} \, c^\dagger_k + \sin\frac{\theta_k}{2} \, c_{-k},\label{Eq6e}
\end{align}
where
 \begin{align}\label{Eq2a}
 \cos\theta_k = \frac{ h- \cos \varphi_k}{\sqrt{h^2 -  2h\cos \varphi_k+1}}, \\
 \sin\theta_k = \frac{ \sin \varphi_k }{\sqrt{h^2 -  2h\cos \varphi_k +1}}.
 \label{Eq2b}
 \end{align}

With help of Eqs. (\ref{Eq6d}) -- (\ref{Eq6e}) we obtain the diagonalized Hamiltonian as a sum of quasiparticles with half-integer quasimomenta,
\begin{align}\label{H2}
{\mathcal H}= \frac{1}{2}\sum_{k}\varepsilon_{0k}  +\sum_{k} \varepsilon_k\Big( a_k^\dagger a_k +\frac{1}{2}\Big) .
\end{align}
where $\varepsilon_{0k}= J (h -\cos \varphi_k)$ and
\begin{align}
  \varepsilon_k = J\sqrt{h^2 -  2h\cos \varphi_k +1}. 
\label{Eps}
\end{align}
Its spectrum contains only states with even number of quasiparticles.

In the momentum representation, the Hamiltonian splits into a sum of independent terms, $  {\mathcal H}= \sum_{k>0}\mathcal H_k$, where each $\mathcal H_k$ acts in the two-dimensional Hilbert space spanned by $|k_0\rangle = |0\rangle_k  |0\rangle_{-k}$ and $|k_1\rangle=|1\rangle_k  |1\rangle_{-k}$.  Here $|0\rangle_k $ is the vacuum state of the mode $c_k$, and $|1\rangle_k  $ is the excited state: $|1\rangle_k =c^\dagger_k |0\rangle_k$. The total wavefunction can be written as, $|\psi(t)\rangle =  \bigotimes_{k>0}|\psi_k(t)\rangle$, where
\begin{align}\label{Eq10}
|\psi_k(t)\rangle=   u_k(t)|k_0\rangle + v_k(t) |k_1\rangle,
\end{align}
and  $|\psi_k\rangle$ satisfies the Bogoliubov-de Gennes equation (in units $\hbar =1$): 
 \begin{align}\label{Eqh1}
i\frac{\partial }{\partial t}|\psi_k\rangle&= \mathcal {H}_k(t)|\psi_k\rangle .
\end{align}

Choosing the basis as, $k_1= \scriptsize\left(
                              \begin{array}{c}
                                1 \\
                                0 \\
                              \end{array}
                            \right)$
and $k_0=\scriptsize\left( \begin{array}{c}
                                0 \\
                                1 \\
                              \end{array}
                            \right)$,
one can show that the Hamiltonian, $ \mathcal {H}_k$, projected on this two-dimensional subspace takes the form,
\begin{align}\label{H1g}
 \mathcal {H}_k  = {\varepsilon_{0k}} {1\hspace{-.125cm}1} +J\left(
\begin{array}{cc}
             h- \cos \varphi_k & \sin \varphi_k \\
             \sin \varphi_k & -h+ \cos \varphi_k  \\
            \end{array}
          \right).
 \end{align}
 For each $k$, there are two eigenstates: 
\begin{align}\label{E2a}
 |u_{+}(k)\rangle =& \left(\begin{array}{c}
                 \cos\frac{\theta_k}{2} \\
                  \sin\frac{\theta_k}{2}
                  \end{array}\right), \\
|u_{-}(k)\rangle = &\left(\begin{array}{c}
               -\sin\frac{\theta_k}{2}\\
                 \cos\frac{\theta_k}{2}
                  \end{array}\right).
                  \label{E2b}
                   \end{align}
 
Since $  {\mathcal H}= \sum_{k>0}\mathcal H_k$, the ground state of the Ising chain can be written as a product of qubit-like states: 
\begin{align}
|\psi_g\rangle =  \bigotimes_{k>0} \Big(\cos\frac{\theta_k}{2}|0\rangle_k  |0\rangle_{-k}  
-\sin\frac{\theta_k}{2}|1\rangle_k  |1\rangle_{-k} \Big).
\end{align}

 For $h\gg 1$, the ground state is paramagnetic with all spins oriented along the $x$ axis, and  from Eq. (\ref{Eq2a}) we obtain $\cos\theta_k \rightarrow 1$ as $h \rightarrow \infty$.  This yields $|u_{-}(k)\rangle \rightarrow \scriptsize 
 \left(\begin{array}{c}
 0 \\
  1 \\
  \end{array}
  \right)$ 
  and 
 $|u_{+}(k)\rangle \rightarrow \scriptsize   \left(\begin{array}{c}
 1 \\
  0 \\
  \end{array}
  \right)$.
On the other hand, when $h\ll 1$ there are two degenerate ferromagnetic ground states with all spins polarized in opposite directions along the $z$-axis. In the thermodynamic limit the system passing through the critical point ends in a superposition of up and down states with finite domains of spins separated by kinks \cite{DJ}.

\section{Quench dynamics }

\subsection{Adiabatic and non-adiabatic evolution}

We consider quantum Ising chain driven by time-dependent Hamiltonian,  ${\mathcal H}(t)= \sum_{k}\mathcal H_k(t)$, where
\begin{widetext}
	\begin{align}\label{H2g}
 \mathcal {H}_k (t) ={\varepsilon_{0k}} (t){1\hspace{-.125cm}1} 
 +J\left(
\begin{array}{cc}
             h(t)- \cos \varphi_k & \sin \varphi_k \\
             \sin \varphi_k & -h(t)+ \cos \varphi_k  \\
            \end{array}
          \right).
\end{align}
\end{widetext}

For a generic quantum system governed by the time-dependent Hamiltonian the adiabatic theorem guarantees that during quantum evolution the system remains in its the ground state, as long as the instantaneous ground state does not become degenerate at any time.
The validity of the adiabatic theorem requires
\begin{equation}\label{QA2}
    \sum_{m\neq n}\bigg|\frac{\langle \psi_m|\partial \mathcal H(t)/\partial
    t|\psi_n\rangle}{(E_m - E_n)^2}\bigg|\ll 1.
\end{equation}

When the quantum processes is related to the quantum phase transitions, the condition of Eq. (\ref{QA2}) can be recast as \cite{SSMO,DC},
\begin{equation}\label{QA3}
  \frac{|\langle \psi_e|\partial{\mathcal H}(t)/\partial
    t|\psi_g\rangle|}{|E_e - E_g|^2} \ll 1,
\end{equation}
where $|\psi_g\rangle$ is the ground state, and $E_e$ is the energy of the first excited state, $|\psi_e\rangle$.  This restriction is violated near the degeneracy in which the QPT occurs.

In the adiabatic basis formed by the instantaneous eigenvectors of the Hamiltonian $H_k$,  the total wavefunction  can be written, as $|\psi\rangle = \otimes_k |\psi_k(t) \rangle $,  where
\begin{align}\label{Eq6a}
 |\psi_k(t) \rangle = & \alpha_k(t) e^{i\int \varepsilon_{0k}(t)dt}  |u_{-}(k,t)\rangle\nonumber \\
& + \beta_k(t)  e^{i\int \varepsilon_{0k}(t)dt} |u_{+}(k,t)\rangle.
\end{align}
From Eqs. (\ref{E2a}) and (\ref{E2b}) it follows that
\begin{align}\label{Eq7z}
\alpha_k(t) =u_k(t)\cos\frac{\theta_k(t)}{2}- v_k(t)\sin\frac{\theta_k(t)}{2}, \\
\beta_k(t) =v_k(t)\cos\frac{\theta_k(t)}{2}+ u_k(t)\sin\frac{\theta_k(t)}{2}.
\end{align}

We define 
\begin{align}
|\Psi_k(t)\rangle = \left (\begin{array}{c}
  \beta_k(t) \\
   \alpha_k(t) 
 \end{array}
\right ).
\end{align}
Next, one can show that the wave function, $|\Psi_k (t)\rangle$, satisfies the Bogoliubov-de Gennes equation
 \begin{align}\label{Eq3A}
i\frac{\partial }{\partial t}|\Psi_k\rangle&= {H}_k(t)|\Psi_k\rangle ,
\end{align}
where
\begin{align}
\label{Eq8}
 {H}_k  = {\varepsilon_{0k}} {1\hspace{-.125cm}1} +\left(
\begin{array}{cc}
          \varepsilon_k & i\dot \theta_k/2\\
           -i\dot \theta_k/2& - \varepsilon_k  \\
            \end{array}
          \right),
\end{align}
and 
\begin{align}
\frac{d\theta_k}{dt} = -\frac{\dot h (t) \sin^2\theta_k(t)}{\sin\varphi_k}
\label{Theta}
\end{align}

Now the requirement of the adiabatic theorem (\ref{QA3}) can be rewritten as,
\begin{align}\label{QA3a}
\max\bigg | \frac{d\theta_k}{dt} \bigg| \ll \min 2\varepsilon_k= 2J\sin\varphi_k.
    \end{align}
Employing (\ref{QA3a}), one can recast Eq. (\ref{Theta}) as,
\begin{align}
|\dot h_c | \ll 2J \sin^2\varphi_k,
\label{Ad1}
\end{align}
where $ \dot h_c = \dot h(t_c) $. Here $t_c$ denotes the moment of time when the magnetic field reached its critical value,  $h_c = 1$.
Further, it is convenient to introduce the following notation:
\begin{align}
\omega_k^2 = \frac{J\sin^2\varphi_k}{|\dot h_c|}.
\label{omega}
\end{align}

Using (\ref{omega}) in Eq. (\ref{Ad1}),  we find that for the given value of $k$ the condition of adiabaticity  can be written as $\omega_k^2 \gg 1$. As follows from  Eq. (\ref{QA3}), for the whole system the condition of adiabaticity can be written as, $\omega^2 \gg 1$, where
\begin{align}
\omega^2\equiv \omega_1  = \frac{J}{|\dot h_c|} \sin^2\bigg( \frac{\pi }{N} \bigg) \gg 1.
\label{omega1}
\end{align}
 For $N\gg 1$ we obtain
\begin{align}
\omega^2 = \frac{\pi^2 J}{|\dot h_c|N^2} \gg 1.
\label{omega2}
\end{align}

Let us assume that the dependence of the magnetic field on time has the form $h=h(t/\tau_0)$. By presenting,
\begin{align}\label{Eq5f}
 \alpha_k(t) = a_k(t)\,e^{i\int_0^t \varepsilon(k,t) dt}, \\
  \beta_k(t) = b_k(t)\,e^{-i\int_0^t \varepsilon(k,t) dt},
  \label{Eq5g}
\end{align}
one can show that, if the evolution begins from the ground state, the coefficients $a_k(t)$ and $b_k(t)$ satisfy the following asymptotic conditions \cite{BMV1,JMPC,JKP,KP,SVG,DG, DAM,DJP,HJP}:
\begin{align}
b_k(t) =  {\mathcal O}\bigg( \exp{\Big(2\tau_0 { \int_0^{z_c} \varepsilon_k(z) dz}\Big)}\bigg),
\label{V2a}
\end{align}
where the critical point, $z_c$, lies on the first Stokes line in the {\em lower} complex line defined as
\begin{equation}\label{St1}
   \Im{ \int_0^{z_c} \varepsilon_k(z) dz} < 0.
\end{equation}
The critical point is determined as a solution of the equation, $\varepsilon_k(z_c)=0$, in the complex plane obtained by analytical continuation, $t \rightarrow z$. 

Employing (\ref{Eps}), we find that for the Ising model the integral in the r.h.s. of Eq. (\ref{V2a}) can be recast as follows:
\begin{align}
\int_0^{z_c }\varepsilon_k(z)dz =\int_0^{e^{-i\varphi_k} } {\sqrt{Z^2 - 2Z\cos\varphi_k +1}}\,\frac{dZ}{Z'},
\label{Eq7a}
\end{align}
where  $Z' = dZ/dz$, and we set $Z=h(z)$.

For given $k$, the probability to stay in the ground state at the end of the evolution is given by
$P^{gs}_k(t) = {|\alpha_k(t)|^2}$. With help of Eq. (\ref{V2a}) we obtain
\begin{align}\label{AEq1}
P^{gs}_k(\infty) =1 - {|\beta_k(t)|^2}\approx 1 - e^{4\tau_0\Im \int_0^{z_c} \varepsilon_k(z)dz}
\end{align}

Since different pairs of quasiparticles ($k,-k$)
evolve independently, the probability to stay
in the ground state for the whole system is the product \cite{DJ}
\begin{align}
P_{gs} = \prod_{k>0}P_{gs}(k).
\label{Eq8g}
\end{align}

For slow evolution one can use the LZ approximation  in Eq. (\ref{Eq7a}), that consists in 
changing   $Z'(z)$ by its value in the critical point, $Z'(z_c)$.  Performing integration  in 
(\ref{Eq7a}) and inserting the result in (\ref{AEq1}), we obtain the Landau Zener formula 
\cite{LL,ZC}
\begin{align}\label{Pq1}
P^{gs}_k \approx1-e^{-\pi \omega_k^2} ,
\end{align}
where $\omega^2_k =(J/{|\dot h_c|}) {\sin^2 \varphi_k}$. This result is valid when $\omega_k^2 \gg1$, that is in agreement with the condition  (\ref{omega1}).  By applying (\ref{Eq8g}) we find that the probability to stay in the ground state for the whole system is 
\begin{align}
P_{gs} = \prod_{k>0}\big(1-e^{-\pi \omega_k^2}\big).
\label{Eq8h}
\end{align}

\subsection{Adiabatic-impulse approximation}

In according to KZ mechanism, the main contribution to the QPT is made in the neighborhood of the 
critical point \cite{KTW,ZHW,ZHW1}. Expanding $h(t)$ near of critical point as $h(t) = 1 + {\dot h}_c (t- 
t_c) $, one can  rewrite (\ref{H2g}) as follows (we omit unessential diagonal contribution, 
${\varepsilon_{0k}}$),
\begin{widetext}
\begin{align}\label{H3g}
\mathcal {H}_k (t) \approx J\left(
\begin{array}{cc}
             {\dot h}_c (t- t_c)+ 2\sin^2(\varphi_k/2)& \sin \varphi_k \\
             \sin \varphi_k & -{\dot h}_c (t- t_c) - 2\sin^2(\varphi_k/2) \\
            \end{array}
          \right).
\end{align}
\end{widetext}
It convenient to introduce a new variable, $\tau_k = (J/\kappa)(t- t_c) + {\rm sgn}(\dot h_c)2\kappa \sin^2(\varphi_k/2)$,  where $\kappa=\sqrt{J/|{\dot h}_c|} $. Let us assume that ${\dot h}_c <0$, then the Hamiltonian (\ref{H2g}) takes the form of the LZ model,
\begin{align}\label{H2f}
  \hat{\mathcal {H}}_k =\left(
\begin{array}{cc}
            -\tau_k& \omega_k \\
            \omega_k & \tau_k \\
            \end{array}
          \right),
\end{align}
the coupling strength being $\omega_k =\kappa\sin\varphi_k $. 
Qualitatively, the dynamics of the LZ system can be described by using so-called the {\em adiabatic-impulse} (AI) approximation \cite{DZ1,ZDZ,DB}.
The AI-approximation assumes that the whole evolution can be divided in three parts, and up to the phase factor the wave function, $|\psi_k(t)\rangle$, approximately can be described as
\begin{align*}
   \tau_k \in [-\infty, - \hat \tau_k ) :  & & |\psi_k(\tau_k )\rangle\approx |u_{-}(k,\tau_k )\rangle  \\
   \tau_k  \in [ - \hat \tau_k,\hat \tau_k ] : & & |\psi_k(\tau_k )\rangle\approx |u_{-}(k,-\hat \tau_k )\rangle   \\
    \tau_k  \in (  \hat \tau_k , +\infty] :  & & |\langle \psi_k(\tau _k)|u_{-}(k,\tau )\rangle|^2 = \rm const
\end{align*}
where the time $\hat \tau_k$, introduced by Zurek \cite{ZHW}, is called the {\em freeze-out time} and define the instant when behaviour of  the system changes from the adiabatic regime to an impulse one where its state is effectively frozen and then back from the impulse regime to the adiabatic one.

If the evolution starts at moment $\tau_i \ll -\hat \tau_k$ from the ground state, the equation for determining $\hat \tau_k$ can be written as  $\pi\hat \tau_k = 1/\rm \varepsilon_k(\hat \tau_k)$ (for details of calculation, see reference \cite{DZ1}). Its solution is given by
\begin{align}\label{LZ1}
\hat \tau_k = \frac{\omega_k}{\sqrt{2}} \sqrt{\sqrt{1+ \frac{4}{\pi^2\omega_k^4}}-1}.
\end{align}

In the AI approximation the probability, $P^k_{ex}$, of exciting mode $k$ at $\tau_f \gg \hat \tau_k$ can be calculated as follows \cite{DZ1,DB}:
\begin{widetext}
	\begin{align}\label{P3}
P^k_{ex} \approx P^k_{AI} = |\langle u_{+}(k,\hat \tau_k)|u_{-}(k,-\hat\tau_k)\rangle|^2
 = \frac{\hat\tau_k^2}{\omega_k^2 +\hat\tau_k^2}.
\end{align}
\end{widetext}
Substituting $\hat \tau_k$ from (\ref{LZ1}), we obtain
\begin{align}\label{P3a}
    P^k_{ex} = \frac{2}{x_k^2 +x_k\sqrt{x_k^2 +4} + 2},
\end{align}
where $x_k= \pi\omega_k^2$. For $\omega_k^2\ll 1$, from Eq. (\ref{P3a}) it follows $P^k_{AI} \approx 1 - \pi \omega_k^2$. In the first order this coincides with the result predicted by exact LZ formula: $P^k_{ex} = e ^{-\pi \omega_k^2}$. For the adiabatic evolution, $\omega_k^2 \gg 1$, we obtain $P^k_{AI} \approx 1/ (\pi^2 \omega_k^4)$. 

Employing  (\ref{Eq8g}) we find that in the AI-approximation the probability to stay
in the ground state for the whole system is given by
\begin{align}
P_{gs} = \prod_{k>0}{\frac{x_k^2 +x_k\sqrt{x_k^2 +4} }{x_k^2 +x_k\sqrt{x_k^2 +4} + 2} }.
\label{EqAI2}
\end{align} 

\begin{figure}[tbp]
\begin{center}
\scalebox{0.425}{\includegraphics{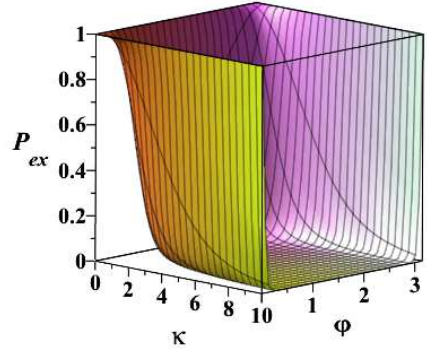}}
\end{center}
\caption{(Color online) Probability of finding the system in the excited state as a function of $\kappa=\sqrt{J/|{\dot h}_c|} $ and $\varphi$.}
\label{PAI}
\end{figure}

In the thermodynamic limit, the variable $\varphi_k$ becomes continuous, and we obtain 
\begin{align}\label{P3b}
    P_{ex}(\kappa,\varphi) = \frac{2}{x^2 +x\sqrt{x^2 +4} + 2},
\end{align}
where $x= \pi\kappa^2\sin^2\varphi$. In Fig. \ref{PAI} the probability of finding the system in excited state is presented. One can see that for $\kappa \gg 1$ the main contribution to $ P_{ex}(\kappa,\varphi)$ is occurred from the values of $\varphi \approx 0$ and $\varphi\approx \pi$. In other limit, $\kappa \ll 1$, the values of $\varphi \approx \pi/2$ yield the most important contribution to the probability $ P_{ex}(\kappa,\varphi)$.

\section{Quench dynamics under shock-wave load}

\subsection{Semi-finite pulse}

\subsubsection{Transition from paramagnet to ferromagnet}

During its evolution the system does not remain in the ground state at all times. At the critical point, 
the quantum system becomes excited, and its final state is determined by the number of defects. In 
the case of the ferromagnetic Ising chain and for transition: {\em  paramagnet $\rightarrow$ 
ferromagnet}, the system ends in the state such as
\begin{align}
\large\boldsymbol |\dots \uparrow\uparrow\uparrow\uparrow 
\downarrow\downarrow\downarrow\downarrow\dots
\downarrow\downarrow\downarrow\uparrow\uparrow\uparrow
\dots\uparrow\uparrow \downarrow
 \downarrow \dots \rangle \nonumber
\end{align}
with neighboring spins polarized in the same directions along the $z$-axis and separated by kinks 
(defects)) in which the polarization of spins has the opposite orientation.

\begin{figure}[tbp]
\scalebox{0.325}{\includegraphics{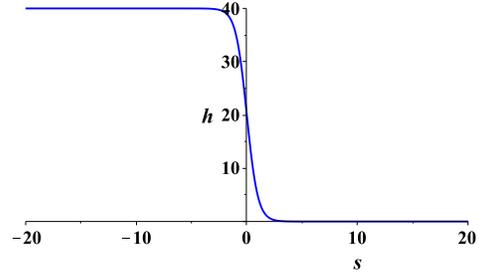}}
\caption{(Color online) Shape of semi-finite pulse: $h$ vs $s=t/\tau_0$ ($h_0 = 20$).}
\label{GP1}
\end{figure}

We specify the magnetic field  as a semi-infinite pulse with the shape determined by (Fig. \ref{GP1})
\begin{align} 
h= h_0(1-\tanh(  t/\tau_0)).
\label{Eq20}
\end{align}

The height of the pulse is given by $h_m = 2h_0$. 

At the critical point, $h_c=1$, we have
\begin{align}
\dot h_c = -\frac{2}{\tau_0}\Big(1-\frac{1}{2h_0}\Big).
\label{Eqha}
\end{align}

It turns out convenient to introduce a dimensionless time $s=  t/\tau_0$ and to recast the Bogoliubov-de Gennes equation (\ref{Eq3A}) as,
\begin{align}\label{Eq3B}
i\frac{\partial }{\partial s}|\Psi_k(s)\rangle&= {\hat H}_k(s)|\Psi_k(s)\rangle ,
\end{align}
where 
\begin{align}
\label{Eq8a}
 {\hat H}_k  = {\hat \varepsilon}_{0k} {1\hspace{-.125cm}1} +\left(
\begin{array}{cc}
          {\hat\varepsilon}_k & i {\theta^\prime}_k/2\\
           -i {\theta^\prime}_k/2& -{\hat \varepsilon}_k  \\
            \end{array}
          \right).
\end{align}
We set ${\theta^\prime}_k =d{\theta}_k/d s$, ${\hat \varepsilon}_{0k}= \tau_0{\varepsilon}_{0k}$, and ${\hat \varepsilon}_k = \tau_0\varepsilon_k$. 

 To estimate asymptotic behavior of the probability to stay in the ground state we use Eq. (\ref{AEq1}).  The result is
 \begin{align}\label{PGS}
 P_{gs}=\prod_k (1 - e^{-4\tau_0\Im \int_0^{z_c} \varepsilon_k(z)dz})
 \end{align}
Using (\ref{Eq20}) in Eq. (\ref{Eq7a}), we are performing integration to obtain 
\begin{widetext}
	 \begin{align}
4\Im \int_0^{z_c }\varepsilon_k(z)dz = {\pi J\tau_0 }(\sqrt{h_m^2-2h_m\cos\varphi +1} +1-h_m)
 \label{Eq7g}
 \end{align}
\end{widetext}
Then the probability to stay in the ground state at the end of evolution can be written as 
\begin{align}
P_{gs} = \prod_{k>0}\big(1-e^{-\pi \nu_k^2}\big),
\label{Eq8c}
\end{align}
where 
\begin{align}
	\nu^2_k =  {\tau_0 J}(\sqrt{h_m^2-2h_m\cos\varphi +1} +1-h_m).
\end{align}
For $h_m \gg 1$ this yields
\begin{align}
  \nu    _k^2 \approx {2J\tau_0}\sin^2\frac{\varphi_k}{2}.
  \label{Omk}
  \end{align} 
Since $\omega_k$ does not includes $h_m$, in the end of evolution the state of system is insensitive to changing of amplitude. Computation of the parameter of adiabaticity, $\omega_k$, yields
\begin{align}
\omega_k^2 = \frac{J\tau_0 \sin^2\varphi_k}{2}.
\label{Ad}
\end{align}
For long wavelength modes with $\varphi_k \ll \pi/4$ we obtain $\nu_k \approx \omega_k$.

\begin{figure}[tbp]
\begin{center}
\scalebox{0.325}{\includegraphics{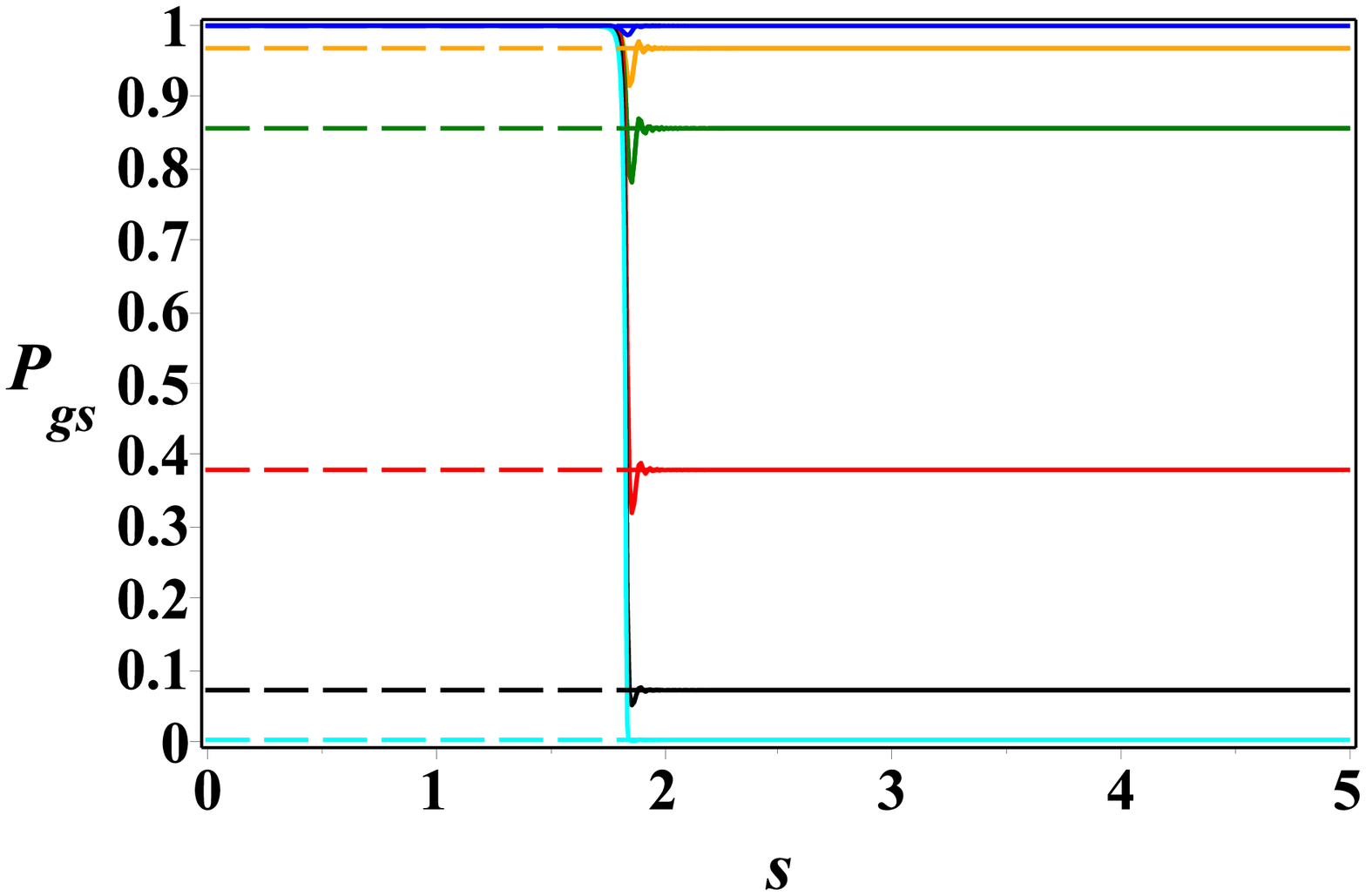}}
(a)
\scalebox{0.325}{\includegraphics{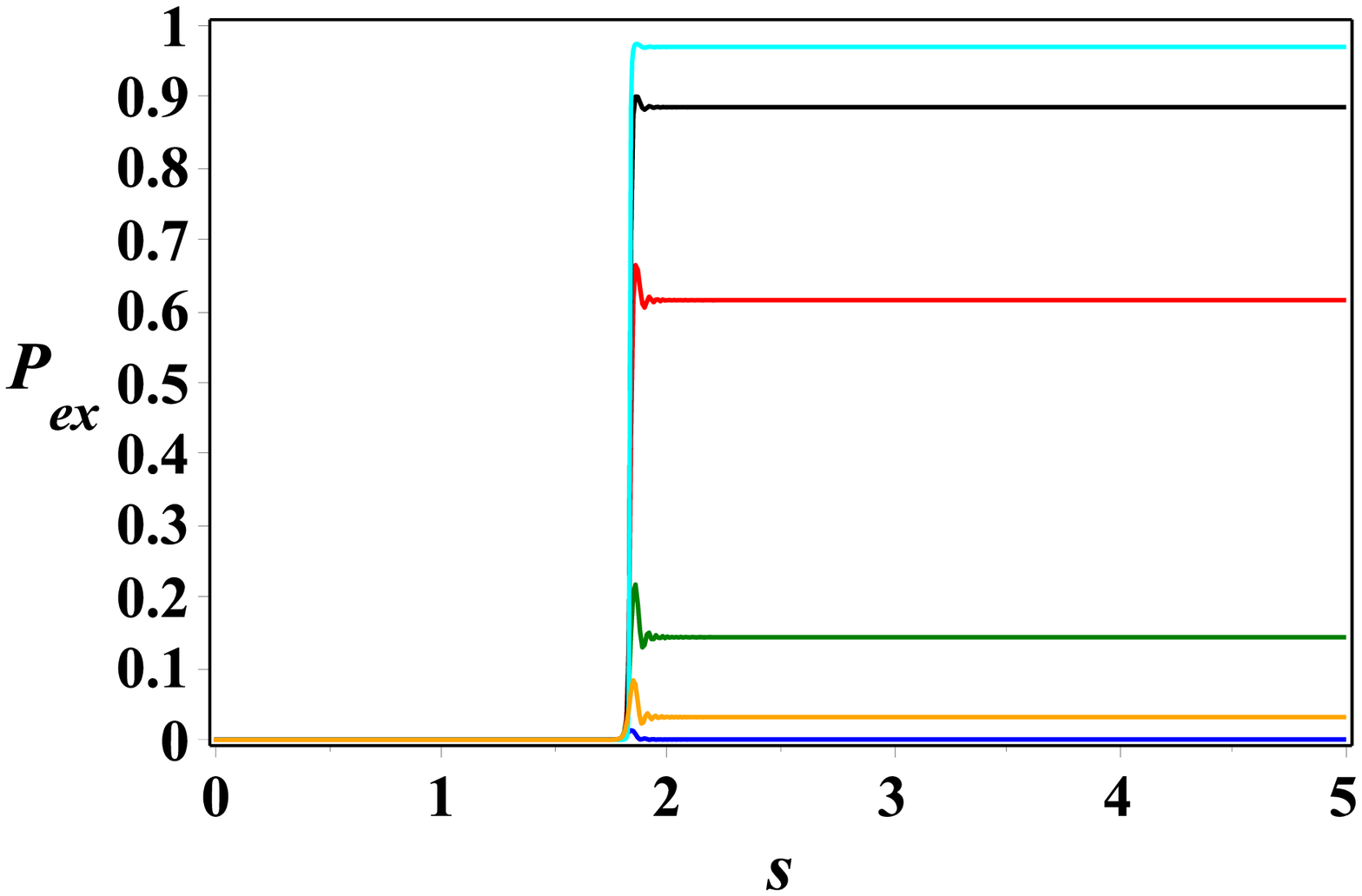}}
(b)
\end{center}
\caption{(Color online) (a) The probability, $P_{gs}$, to stay in the ground state as a function of the dimensionless time $s=t/\tau$.  (b) The probability $P_{ex}$ of first excited state ($k=1$) vs $s$.  Parameters used: $\tau = 500$, ${h}_0 =20$. Blue line ($N=32$), orange line ($N=48$), green line ($N=64$), red line ($N=128$), black line ($N=256$), cyan line ($N=512$). Dashed lines correspond to the asymptotic formula (\ref{Eq8c}).  }
\label{P2}
\end{figure}

\begin{figure}[tbp]
\scalebox{0.325}{\includegraphics{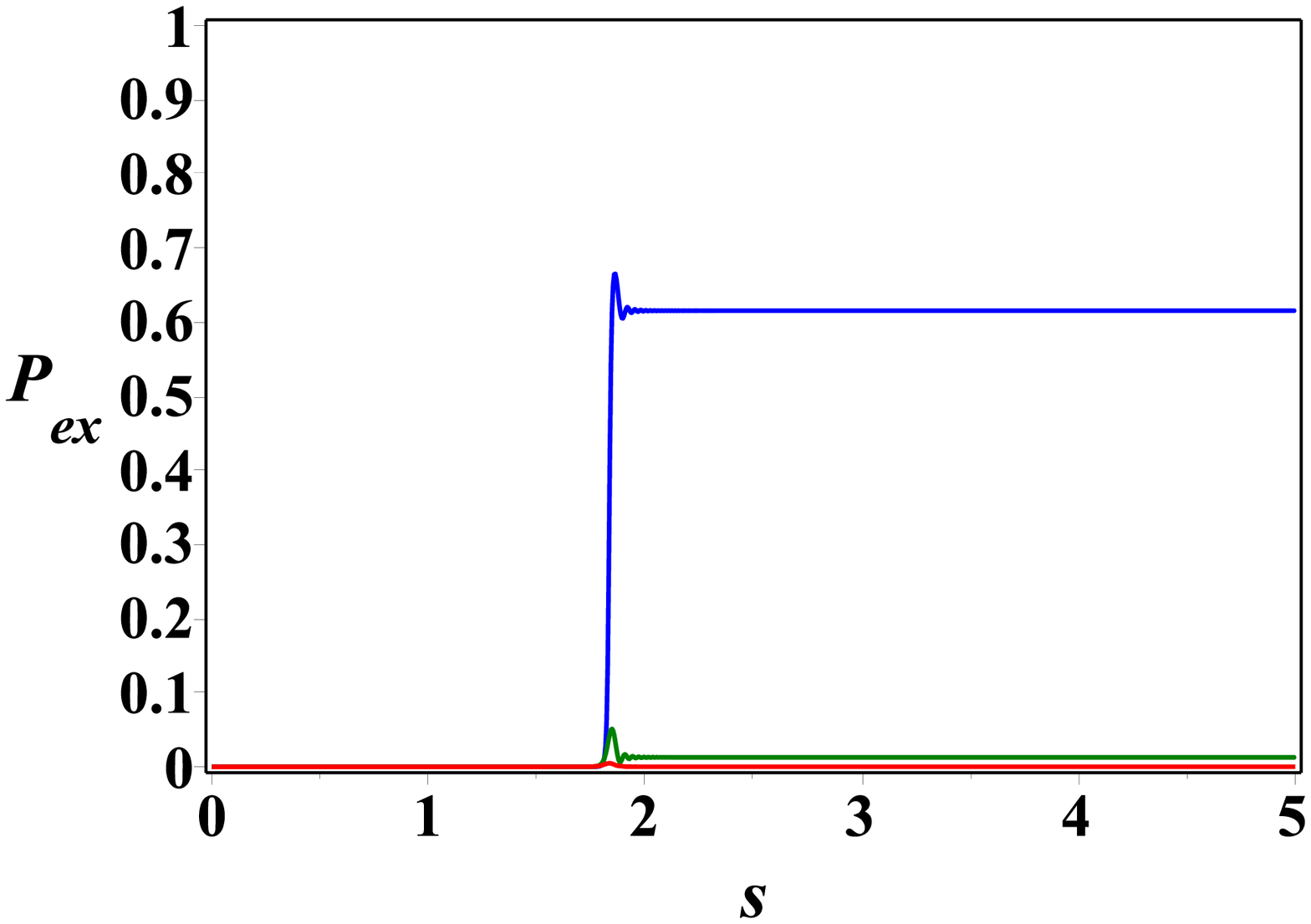}}
(a)
\scalebox{0.325}{\includegraphics{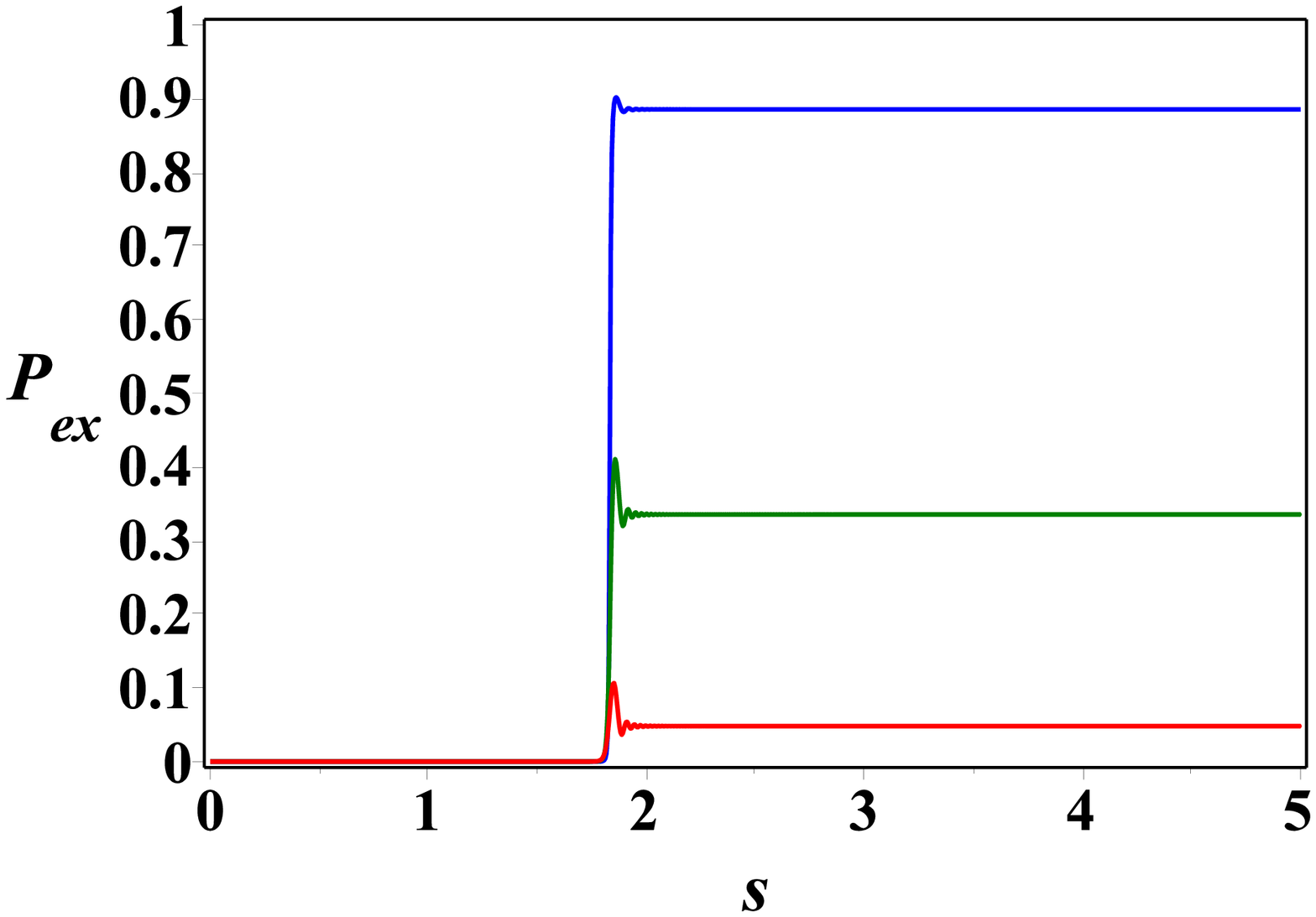}}
(b)
\caption{(Color online)  The probability $P_{ex}$ of first excited states ($k=1,2,3$) as a function of the dimensionless time $s$. Blue line ($k=1$),  green line ($k=2$), red line ($k=3$). (a) $N=128$. (b) $N=256$.  .}
\label{P2b}
\end{figure}

Our theoretical predictions are confirmed by numerical calculations performed for $N=32,48,64,128,256,512$ spins (See Figs. \ref{P2}, \ref{P2b}.). We assume that initially the system was in the ground (paramagnet) state. Choice of parameters: $J=1$, $\tau_0 = 500$, ${h}_0 =20$.  Solid lines present the results of the numerical simulations and dashed lines correspond to the asymptotic formula (\ref{Eq8c}). One can observe that while short wavelength excitations are essential at the critical point, at the end of evolution their contribution to the transition probability from the ground state to the first excited state is negligible. The results presented in Fig. \ref{P2}a show that the asymptotic formula (\ref{Eq8c}) is in good agreement with the results of numerical simulations. 

The estimation of the adiabaticity parameter  $\omega$ (see Eq.(\ref {omega1})) yields: $\omega =2.4$ ($N=32$), $\omega =1.07$ ($N=48$), $\omega =0.6$ ($N=64$),  $\omega =0.15$ ($N=128$), $\omega =0.04$ ($N=256$), $\omega =0.01$ ($N=512$). As expected, with decreasing of $\omega $ the  probability to stay in the ground state decreases as well. This implies that at the end of evolution the quantum system does not remains in the ground state and its  final state is the superposition of blocks with the spins oriented up/down, separated by walls (kinks).

 \subsubsection{Transition from ferromagnet to paramagnet }
  
\begin{figure}[tbp]
\scalebox{0.325}{\includegraphics{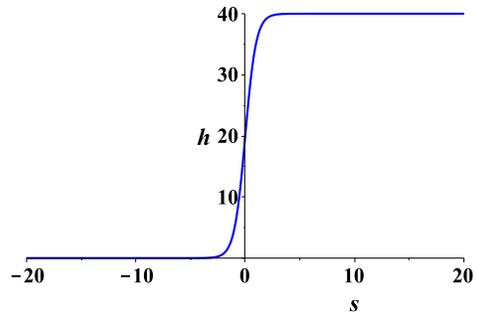}}
\caption{(Color online) Shape of pulse, $h(s)$, as a function of dimensionless time $s=t/\tau_0$ ($h_0 =20$). }
\label{GP1a}
\end{figure}

To describe transition from ferromagnet to paramagnet, we specify the magnetic field  as a semi-infinite pulse with the shape determined by (See Fig. \ref{GP1a}.)
\begin{align} 
 h={ h}_0\big(1+\tanh( t/\tau)\big),
\label{EqP2}
\end{align}
height of the pulse being $h_m = 2h_0$. 

\begin{figure}[tbp]
\begin{center}
\scalebox{0.325}{\includegraphics{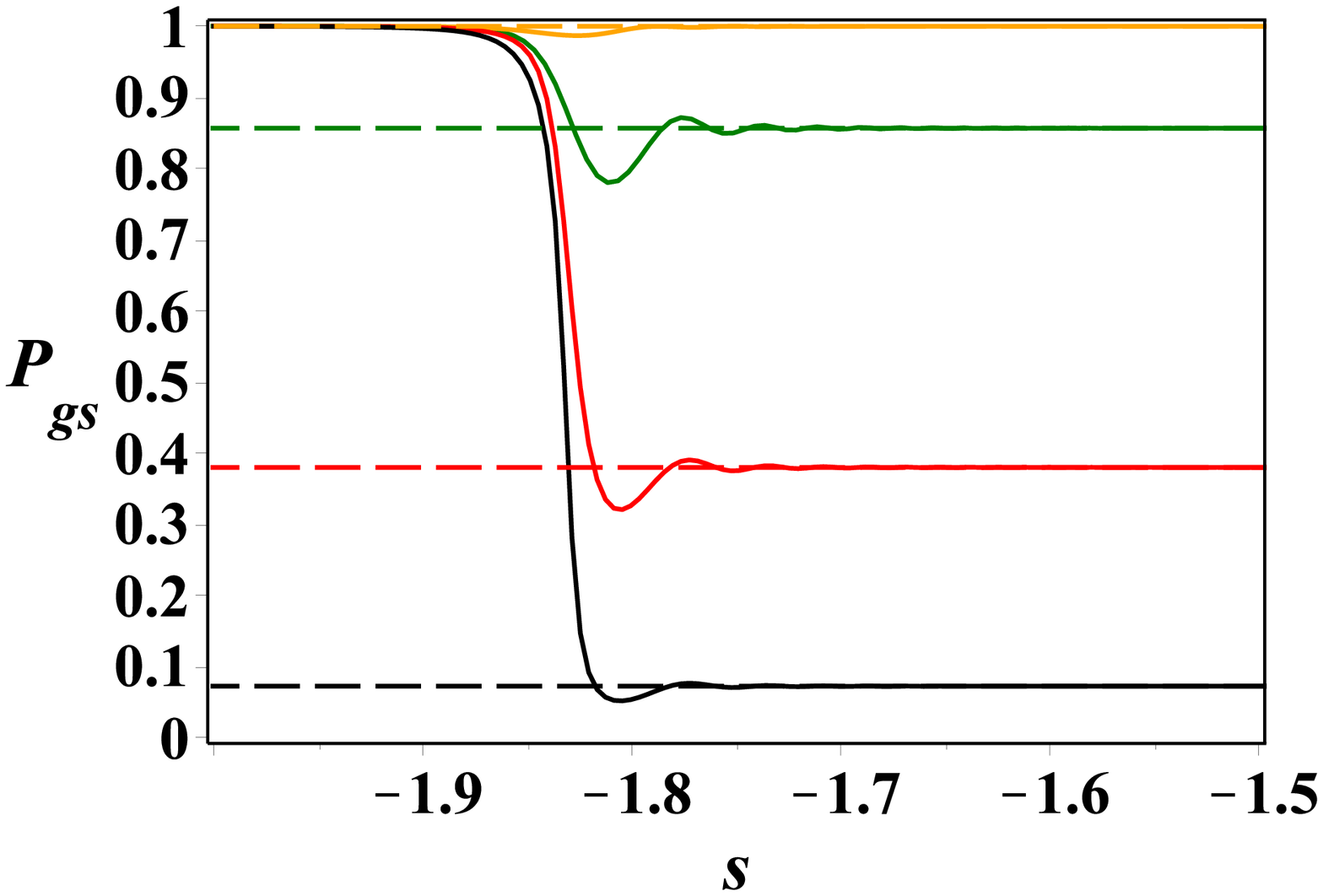}}
(a)
\scalebox{0.325}{\includegraphics{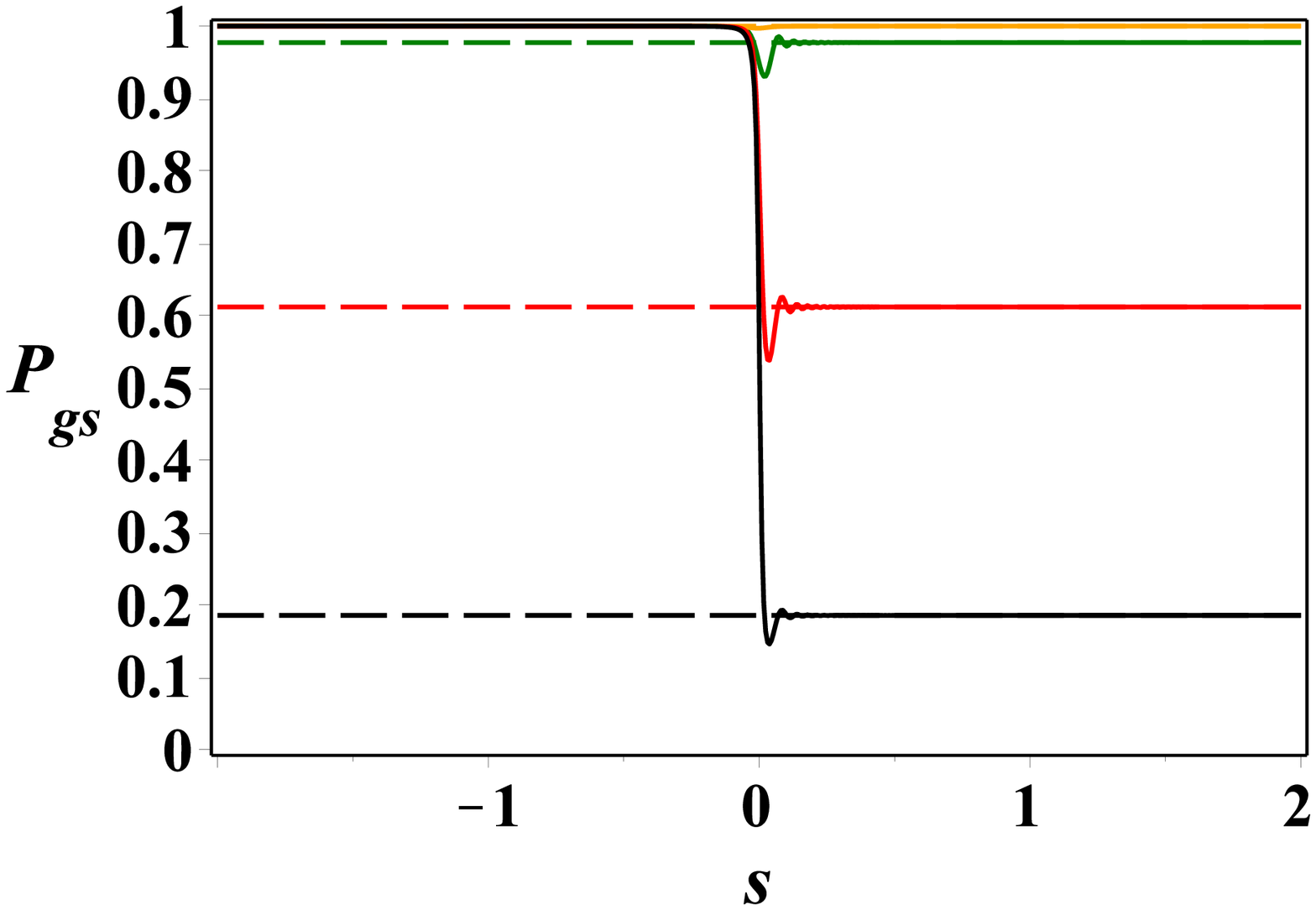}}
(b)
\end{center}
\caption{(Color online) Ferromagnet $\rightarrow$ paramagnet transition ($\tau_0 = 500$). The probability, $P_{gs}$, to stay in the ground state as a function of the dimensionless time $s=t/\tau_0$. Orange line ($N=32$), green line ($N=64$), red line ($N=128$), black line ($N=256$). Dashed lines correspond to the asymptotic formula (\ref{Eq13}). (a) ${h}_0 =20$.  (b) ${h}_0 =1$. }
\label{P2a}
\end{figure}
 
 We start with the initial ferromagnet ground state.  Near the critical point, the quantum system becomes excited, and its final state (for $h_0 \gg 1$) is determined by the number of flipped spins.  
 
 We find that aymptotic behavior of the probability to stay in the ground state is given by
\begin{align}\label{Eq13}
 P_{gs}(\infty) =\prod_k (1 - e^{-\pi\nu_k^2}),
\end{align}
where
  \begin{align}
 \nu^2_k = J {\tau_0}\big(\sqrt{h_m^2-2h_m\cos\varphi_k +1}+1- h_m\big).
  \label{Eq14a}
  \end{align}
When $h_m \gg 1$ one can approximate (\ref{Eq14a}) as follows:
\begin{align}
 \nu^2_k \approx 2J {\tau_0}\sin^2 \frac{\varphi_k}{2}.
 \label{Eq14b}
\end{align}

In Fig. \ref{P2a} we present the results of  numerical simulations performed for 
$N=32,48,64,128,256$ spins. Choice of parameters: $J=1$, $\tau_0 = 500$, ${h}_0 =1,20$.  Solid 
lines present the results of the numerical simulations and dashed lines correspond to the asymptotic 
formula (\ref{Eq8}).  We assume that initially the system was in the ground (ferromagnet) state. One 
can observe that while short wavelength excitations are essential at the critical point, at the end of 
evolution their contribution to the transition probability from the ground state to the first excited 
state is negligible. The results presented in Fig.\ref{P2a}a show that the asymptotic formula 
(\ref{Eq13}) is in good agreement with the results of numerical simulations. 

\subsubsection{LZ formula and AI-approximation}

In this section we compare LZ-formula and  AI-approximation with the results of the numerical simulations for semi-finite pulse. We assume that initially the system was in the ground state, then the probability to stay in the ground state in the end of evolution for the whole system can be written as,
\begin{align}\label{EqLZa}
 {\rm LZ}: \;P^{LZ}_{gs} = &\prod_{k>0}(1 - e^{-x^2_k} ), \\
{\rm AI}: \;
P^{AI}_{gs} =&\prod_{k>0}{\frac{x_k^2 +x_k\sqrt{x_k^2 +4} }{x_k^2 +x_k\sqrt{x_k^2 +4} + 2} }, \; 
\label{EqAI2b}
\end{align} 
 where $x_k= \pi\omega_k^2$, the parameter of adiabaticity being $\omega_k = \sqrt{J/|{\dot h}_c|}\sin\varphi_k$. For the semi-finite pulse introduced in the Secs. 4.1.2. and 4.1.3, we obtain
 \begin{align}
 \omega_k= \sqrt{\frac{\tau_0 h_0 J}{|2h_0 - 1|}}\sin\varphi_k .
  \end{align}

\begin{figure}[tbp]
\begin{center}
\scalebox{0.325}{\includegraphics{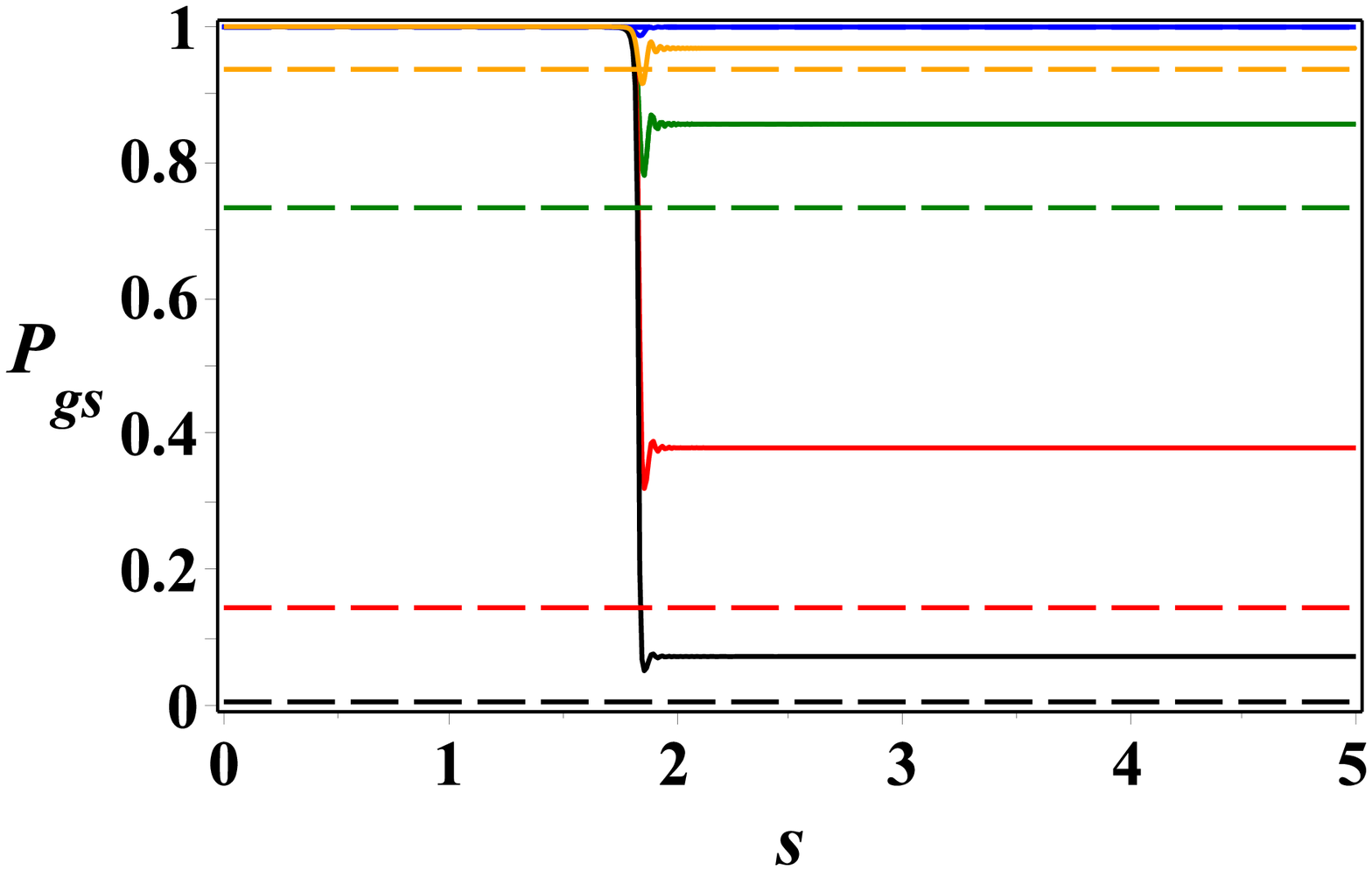}}
(a)
\scalebox{0.33}{\includegraphics{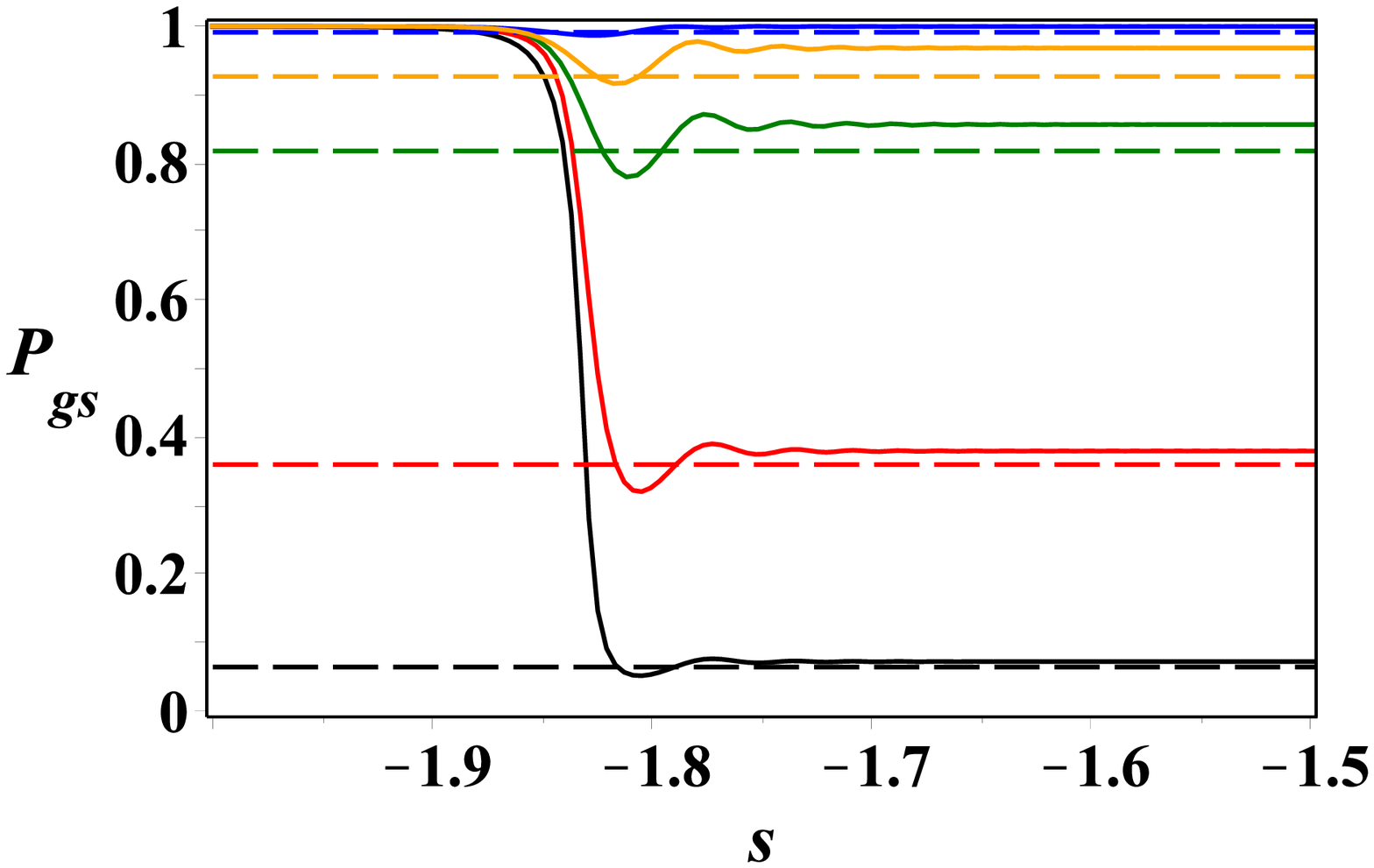}}
(b)
\end{center}
\caption{(Color online)   The probability, $P_{gs}$, to stay in the ground state as a function of the dimensionless time $s=t/\tau_0$. Blue line ($N=32$), orange line ($N=48$), green line ($N=64$), red line ($N=128$), black line ($N=256$). (a) Paramagnet $\rightarrow$ ferromagnet transition. Dashed lines correspond to the LZ-formula (\ref{EqLZa}). (b) Ferromagnet $\rightarrow$ paramagnet transition. Dashed lines correspond to the AI-approximation (\ref{EqAI2b}). Choice of parameters: $\tau_0 = 500$, ${h}_0 =20$. }
\label{FPF}
\end{figure}
\begin{figure}[tbp]
\begin{center}
\scalebox{0.3}{\includegraphics{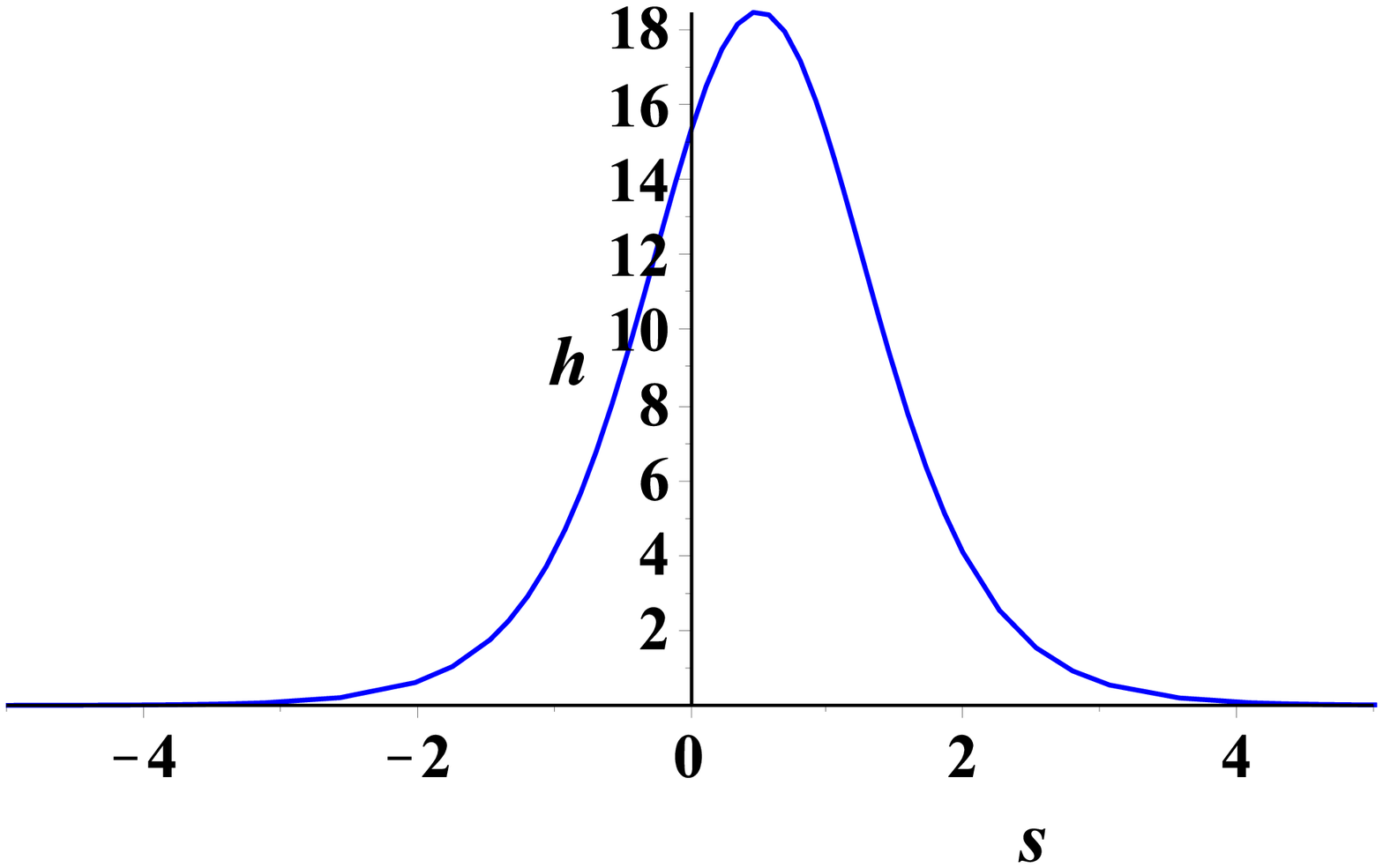}}
(a)
\scalebox{0.325}{\includegraphics{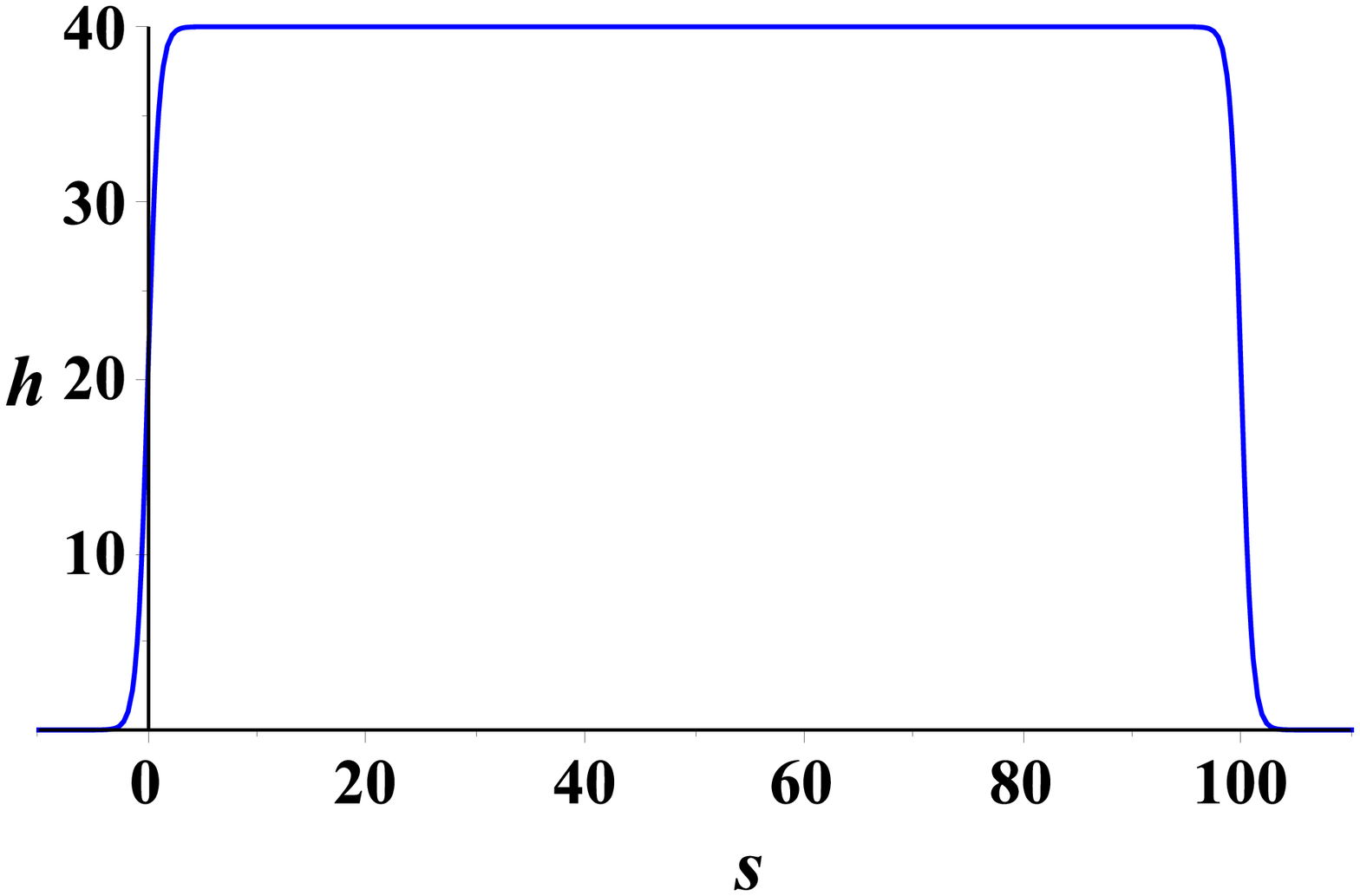}}
(b)
\end{center}
\caption{(Color online) Finite pulse, $h(s)$ as a function of dimensionless time $s=t/\tau_0$ ($h_0 =20$). (a) $\delta   = 1$. (b) $\delta = 100$. }
\label{GP1n}
\end{figure}

In Fig. \ref{FPF}  we compare  the predictions of LZ-formula and AI-approximation  with the results of  numerical simulations performed for $N=32,48,64,128,256$ spins.  Choice of parameters: $J=1$, $\tau_0 = 500$, ${h}_0 =20$.  Solid lines present the results of the numerical simulations and dashed lines correspond to the asymptotic formulas (\ref{EqLZa}) and (\ref{EqAI2b}). We assume that initially the system was in the ground state.  In Fig. \ref{FPF}a the results of the paramagnet $\rightarrow$ ferromagnet transition are presented. As expected, when  the parameter of adiabaticity $\omega^2_k \gg 1$, the LZ-formula is in good agreement with the numerical results (blue and orange curves). In Fig. \ref{FPF}b we compare the AI-approximation with the results of the numerical simulations for ferromagnet $\rightarrow$  paramagnet transition. One can observe that  AI-approximation is good enough for $\omega^2_k \ll 1$ and $\omega^2_k \gg 1$. 
\begin{figure}[tbp]
\begin{center}
\scalebox{0.325}{\includegraphics{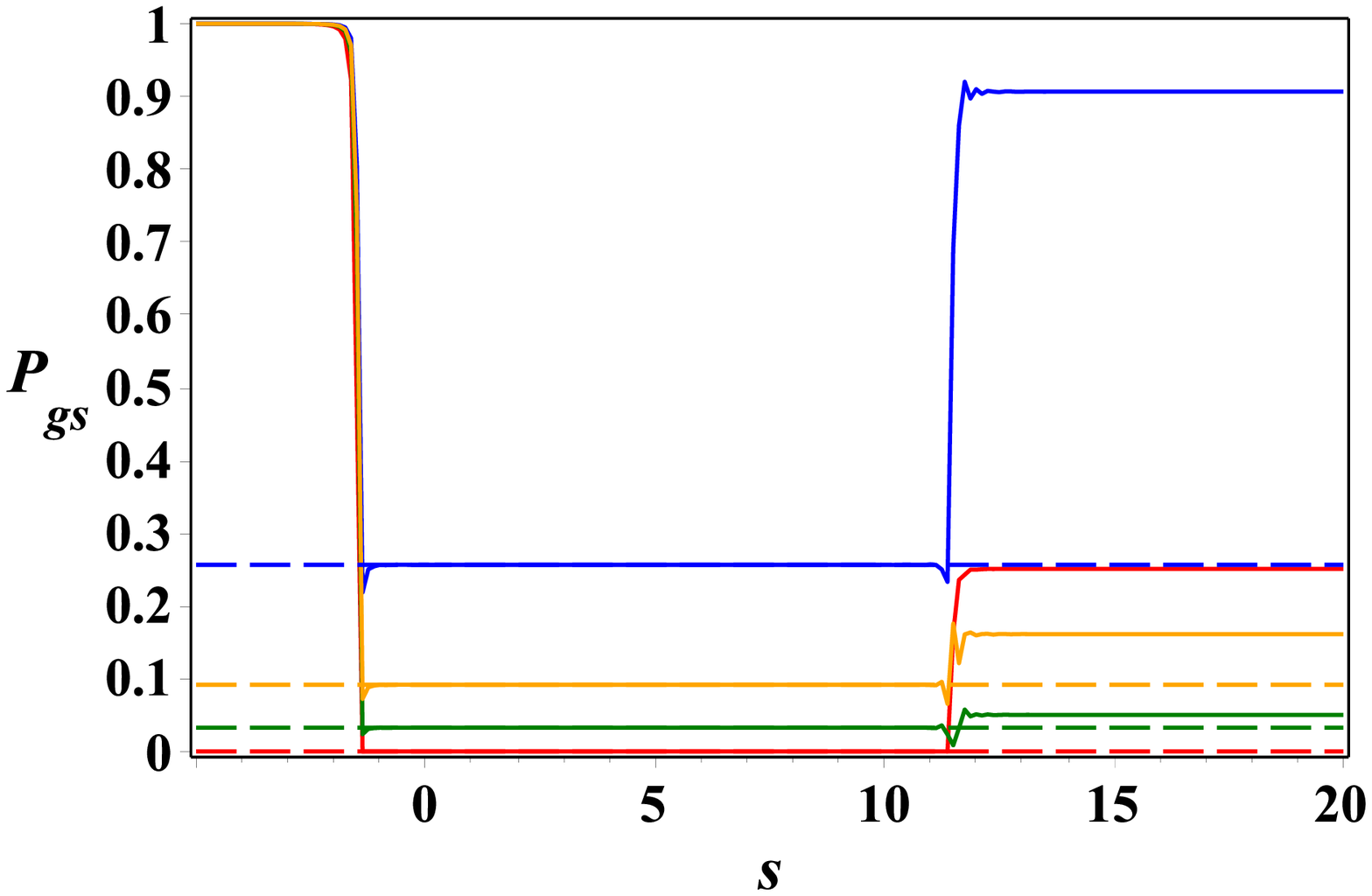}}
(a)
\scalebox{0.325}{\includegraphics{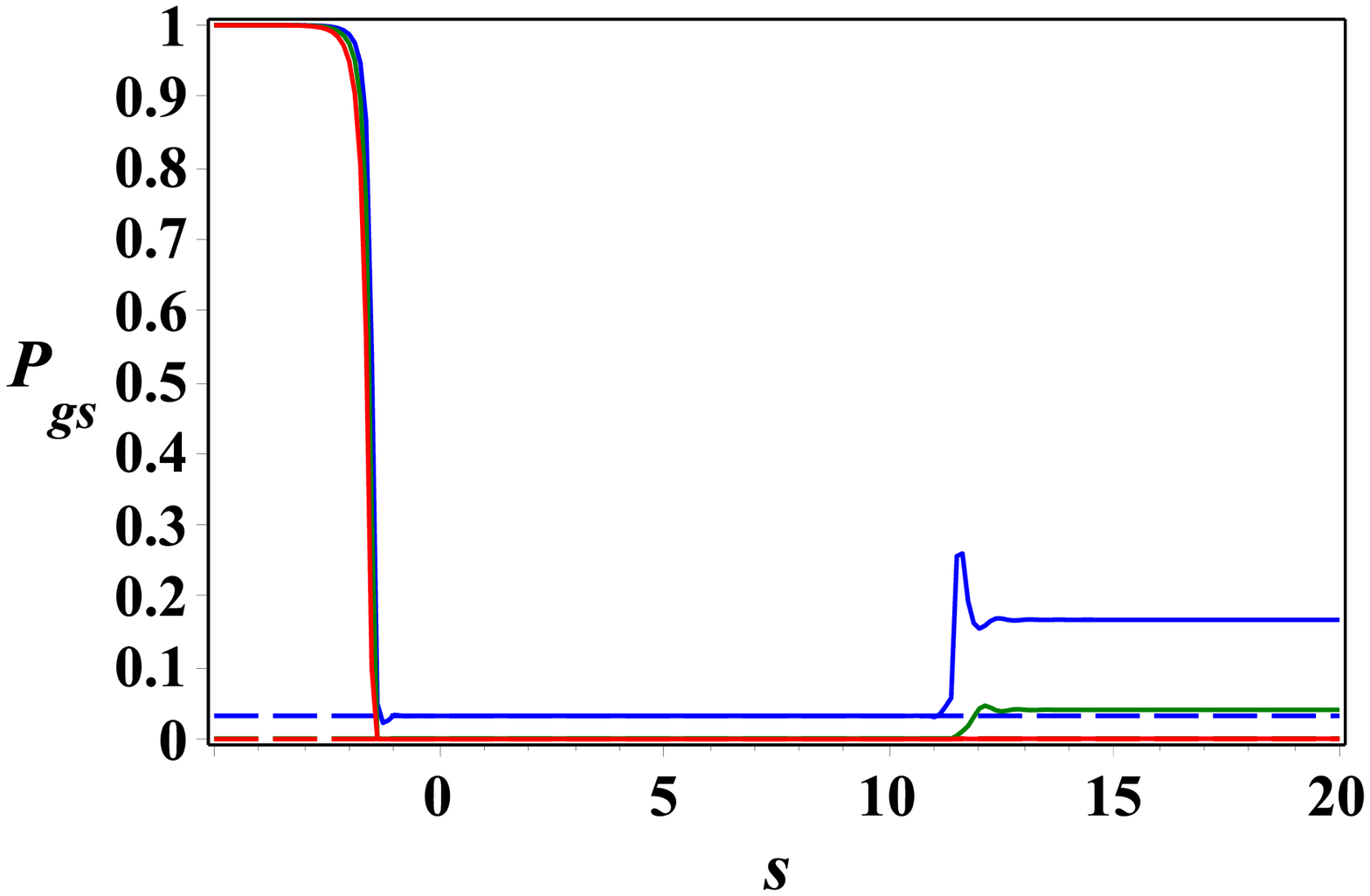}}
(b)
\end{center}
\caption{(Color online)  The probability, $P_{gs}$, to stay in the ground state as a function of the dimensionless time $s=t/\tau_0$ (${h}_0 =10$, $\delta =10$). (a) $\tau_0 = 20$. (b) $\tau_0 = 5$. Blue line ($N=32$), orange line ($N=48$), green line ($N=64$), red line ($N=128$). Dashed lines correspond to the asymptotic formula (\ref{Eq13a}).  }
\label{FP2}
\end{figure}

\subsection{Pulse of finite length}

We consider a magnetic field as a pulse of finite length with the shape determined by 
\begin{align} \label{Eq6_a}
{h}(\tau)=   h_0\big (\tanh( t/\tau_0)-\tanh(t/\tau_0-\delta)\big),
\end{align}
the pulse length being $\Delta =\delta\tau_0  $, and its height is given by $h_m = 2h_0\tanh(\delta/2)$ (see Fig. \ref{GP1n}).

We assume that initially the system was in the ferromagnet ground state. Near the first critical point, the quantum system becomes excited, and after crossing the critical point its  state (for $h_0 \gg 1$) is determined by the number of flipped spins. 

\begin{figure}[tbp]
\begin{center}
\scalebox{0.325}{\includegraphics{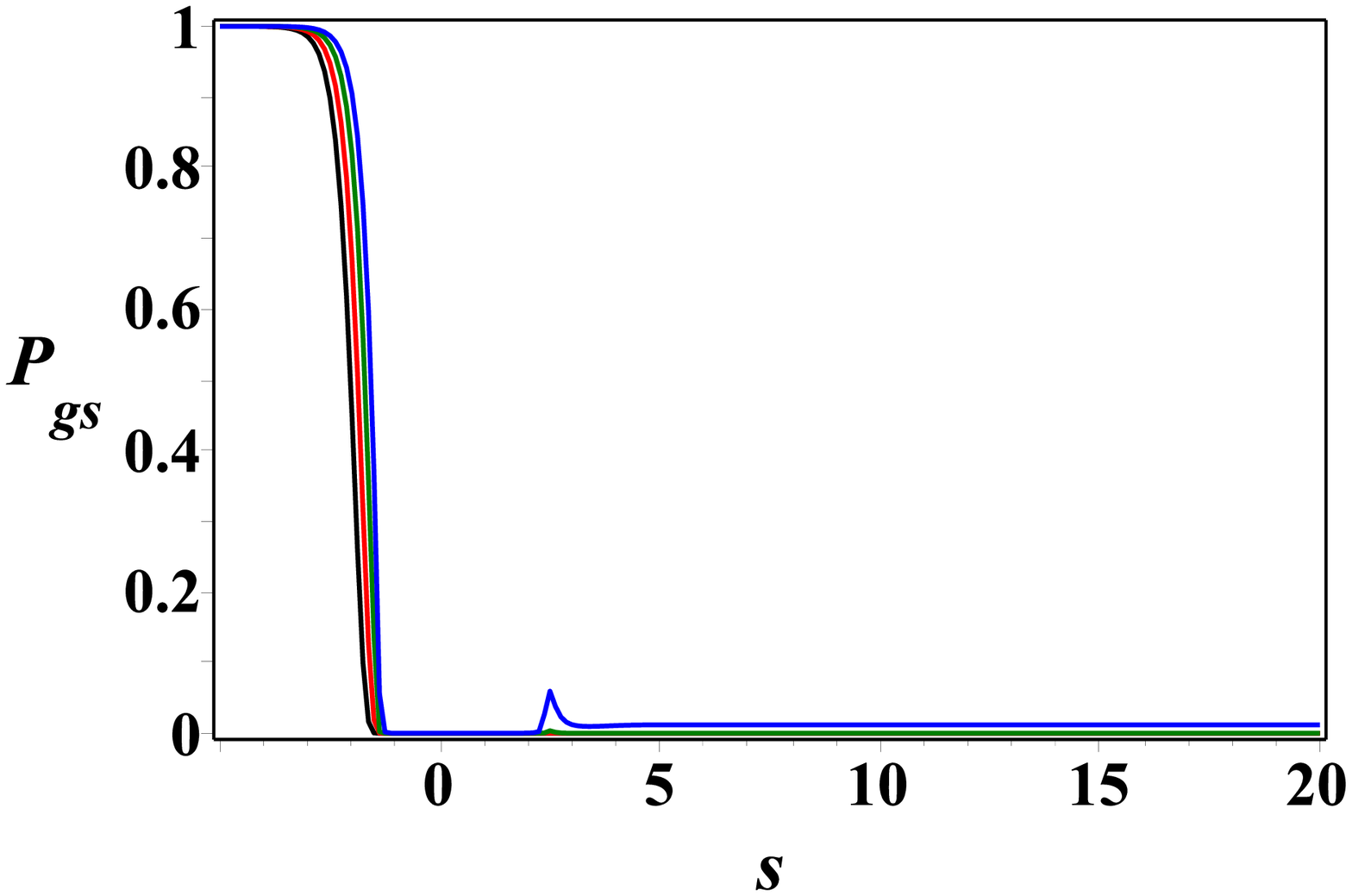}}
(a)
\scalebox{0.325}{\includegraphics{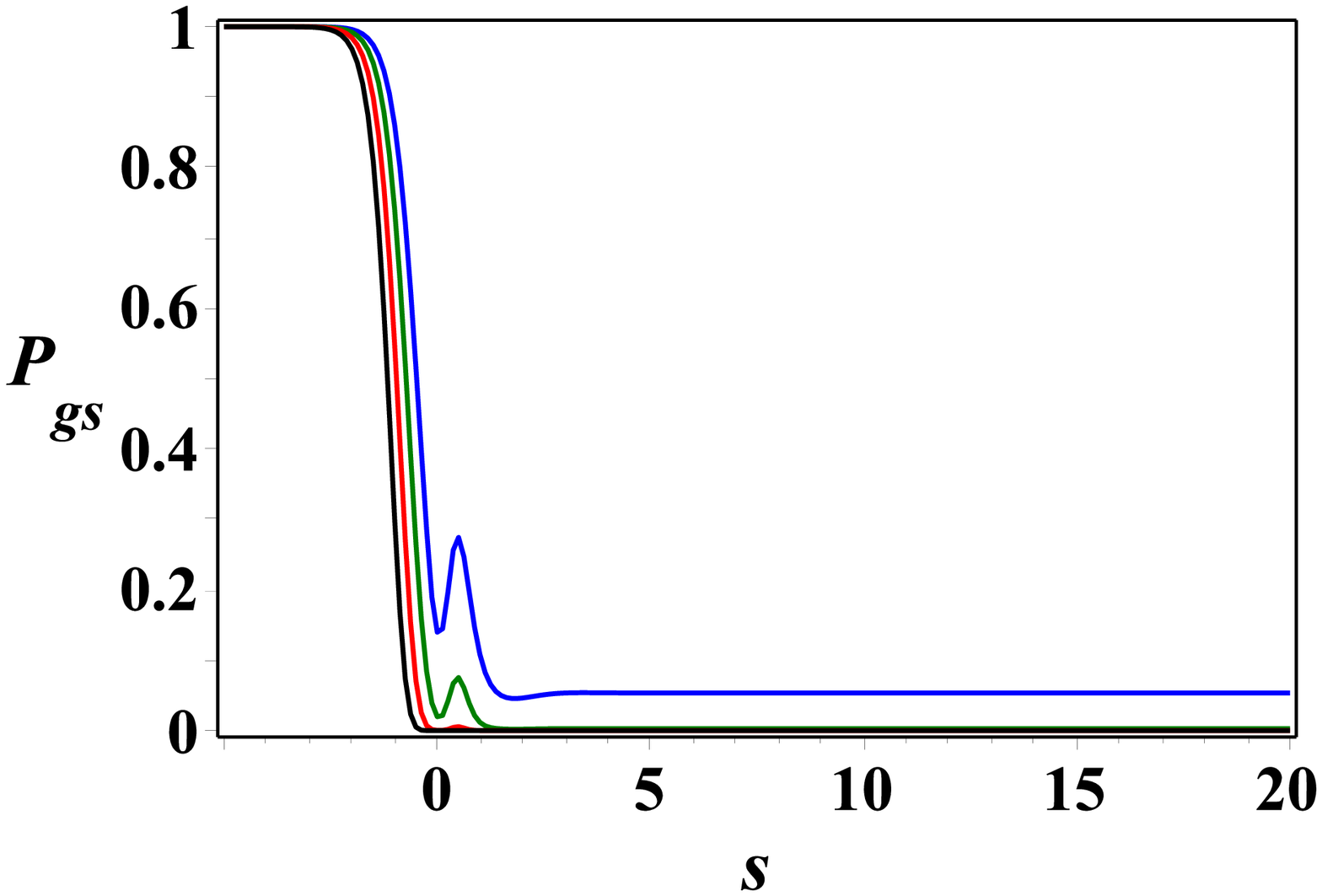}}
(b)
\end{center}
\caption{(Color online) Shock wave load. The probability, $P_{gs}$, to stay in the ground state as a function of the dimensionless time $s=t/\tau_0$ (${h}_0 =10$, $\tau_0 = 1$).  (a)  $\delta =1$). (b) $\delta = 0.1$.   Blue line ($N=32$), green line ($N=64$), red line ($N=128$), black line ($N= 256$). }
\label{FP3}
\end{figure}

\begin{figure}[tbp]
\begin{center}
\scalebox{0.325}{\includegraphics{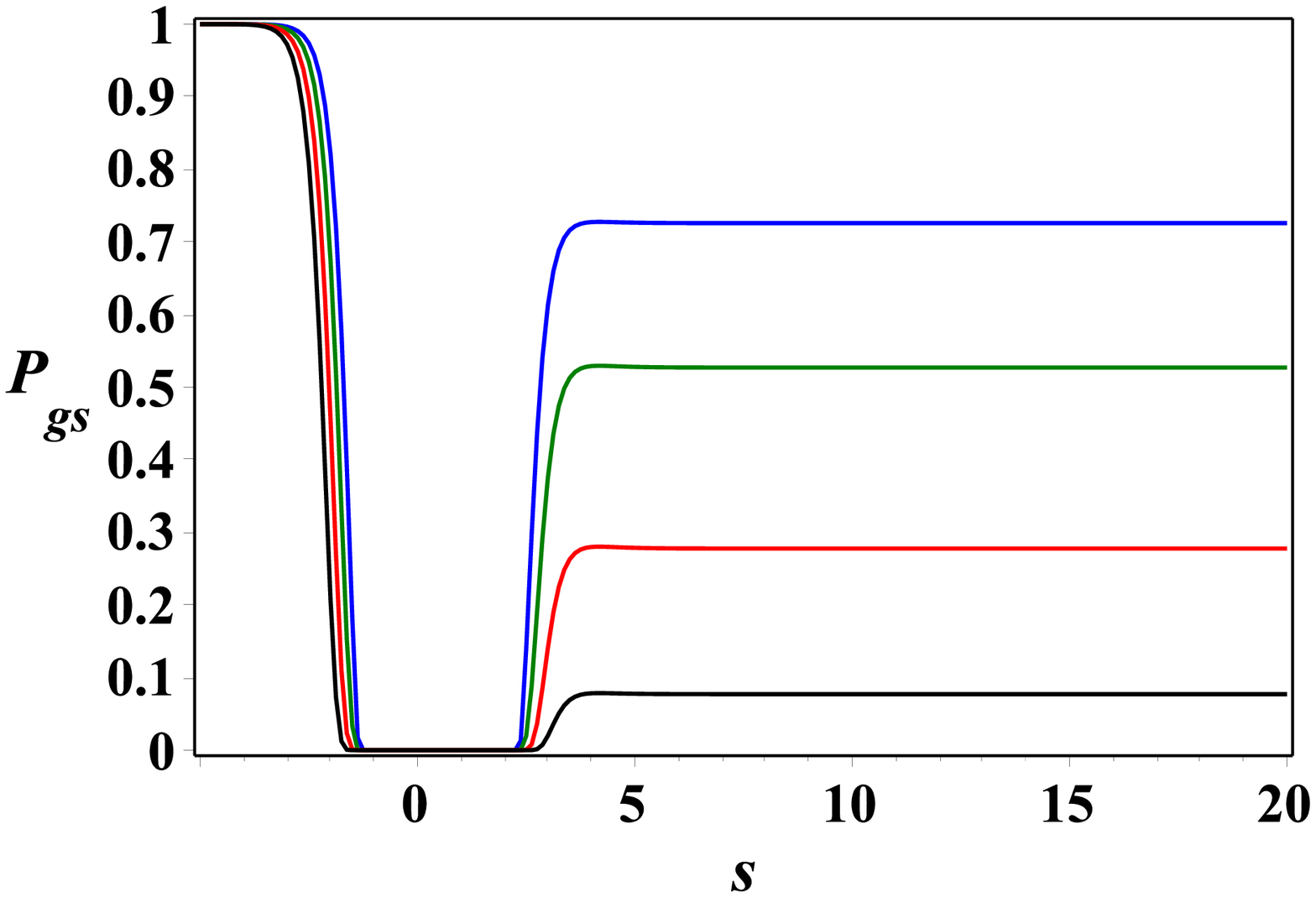}}
(a)
\scalebox{0.32}{\includegraphics{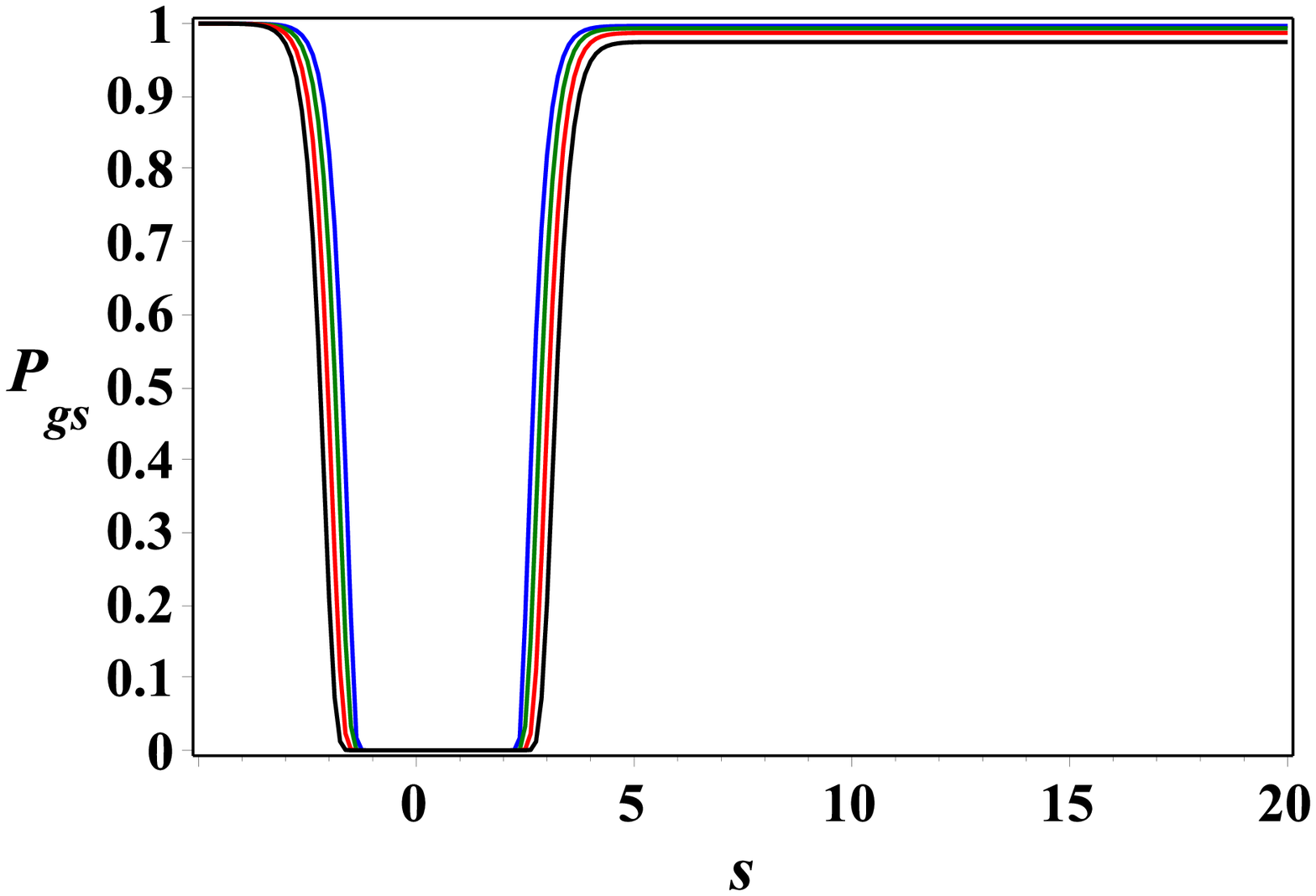}}
(b)
\end{center}
\caption{(Color online) Shock wave load. The probability, $P_{gs}$, to stay in the ground state as a function of the dimensionless time $s=t/\tau_0$ (${h}_0 =10$, $\delta =1$). (a) $\tau_0 = 0.01$. (b) $\tau_0 = 0.001$.  Blue line ($N=32$), green line ($N=64$), red line ($N=128$), black line ($N= 256$).}
\label{FP3a}
\end{figure}

According to the KZ mechanism, the system will be stay in this state up to reaching the second critical point. When the length of the pulse is relatively large,  there exists the intermediate asymptotic for the probability to stay in the ground state. We find that it is given by
\begin{align}\label{Eq13a}
 P_{gs} =\prod_k (1 - e^{-\pi\nu_k^2}),
\end{align}
where
  \begin{align}
 \nu^2_k =  {\tau_0}J\Big(\sqrt{{ h}_m^2- 2{ h}_m\cos\varphi +1} +1 - { h}_m \Big).
  \label{Eq14}
  \end{align}
After crossing the second critical point, the system ends in the state with the domain structure, consisting of domains with neighboring spins polarized in the same directions along the $z$-axis and separated by kinks (defects) in which the polarization of spins has the opposite orientation.

In Figs. \ref{FP2} - \ref{FP3a} we present the results of  numerical simulations performed for $N=32,48,64,128,256$ qubits.  Choice of parameters: $\tau_0 = 0.001, 0.01,1,5,20$, ${h}_0 =10$, $\delta=1,10$. We assume that initially the system was in the ground (ferromagnet) state. Solid lines present the results of the numerical simulations and dashed lines correspond to the asymptotic formula (\ref{Eq13a}). One can observe  good agreement between the intermediate asymptotic given by Eq. (\ref{Eq13a}) and the results of numerical simulations (Fig. \ref{FP2}).

\section{Critical phenomena and defects formation}

During  its evolution the system does not stay always at the ground state at 
all times. At the critical point, the system becomes excited, and its final 
state is determined by the number of defects (kinks). Following \cite{DJ}, we 
define the operator of the number of kinks as,
\begin{align}
 \hat{\mathcal N} = \frac{1}{2}\sum^N_{n=1}\big ( 1- \sigma^z_n \sigma^z_{n+1}\big ) = \sum_{k} a_k^\dagger a_k.
\end{align}
Employing Eqs. (\ref{Eq6d}) - (\ref{Eq6e}), we obtain
\begin{widetext}
\begin{align}\label{Ia}
 \hat{\mathcal N}= \frac{N}{2} +\frac{1}{2}\sum_{k}\Big (\cos\theta_k(c^\dagger_k c_k- c_kc^\dagger_k )
+ \sin \theta_k(c^\dagger_k c^\dagger_{-k} + c_{-k} c_{k} ) \Big),
\end{align}
\end{widetext}

The number of defects is defined by the expectation value of the operator of the number of 
defects, $\mathcal N =\langle  \hat{\mathcal N} \rangle$. The computation yields
\begin{widetext}
\begin{align}
\mathcal N =\frac{N}{2} +\frac{1}{2}\sum_{k}\Big (\cos\theta_k(|v_k|^2 - |u_k|^2)
+ \sin \theta_k(u_k^\ast v_k  + u_k v_k^\ast  ) \Big).
\end{align}
\end{widetext}
In the adiabatic basis this formula takes a more simpler form,
\begin{align}
\mathcal N =\frac{N}{2} -\frac{1}{2}\sum_{k}(|\alpha_k|^2 - |\beta_k|^2).
\end{align}
For the expectation value of the density of defects we obtain,
\begin{align}\label{TL1}
n= \frac{\mathcal N}{N}  = 1- \frac{1}{N}\sum_{k} |\alpha_k|^2.
\end{align}
Denoting the probability to stay in the ground state as $P_{gs}(\varphi_k) \equiv  |\alpha_k|^2$, we rewrite (\ref{TL1}) as
\begin{align}\label{TL1a}
n = 1- \frac{1}{N}\sum_{k} P_{gs}(\varphi_k).
\end{align}
Using the approximated formula (\ref{AEq1}
),  we obtain
\begin{align}\label{D1}
n\approx \frac{1}{N}\sum_{k}e^{4\tau_0\Im \int_0^{z_c} \varepsilon_k(z)dz}
\end{align}
\begin{figure}[tbp]
\begin{center}
\scalebox{0.325}{\includegraphics{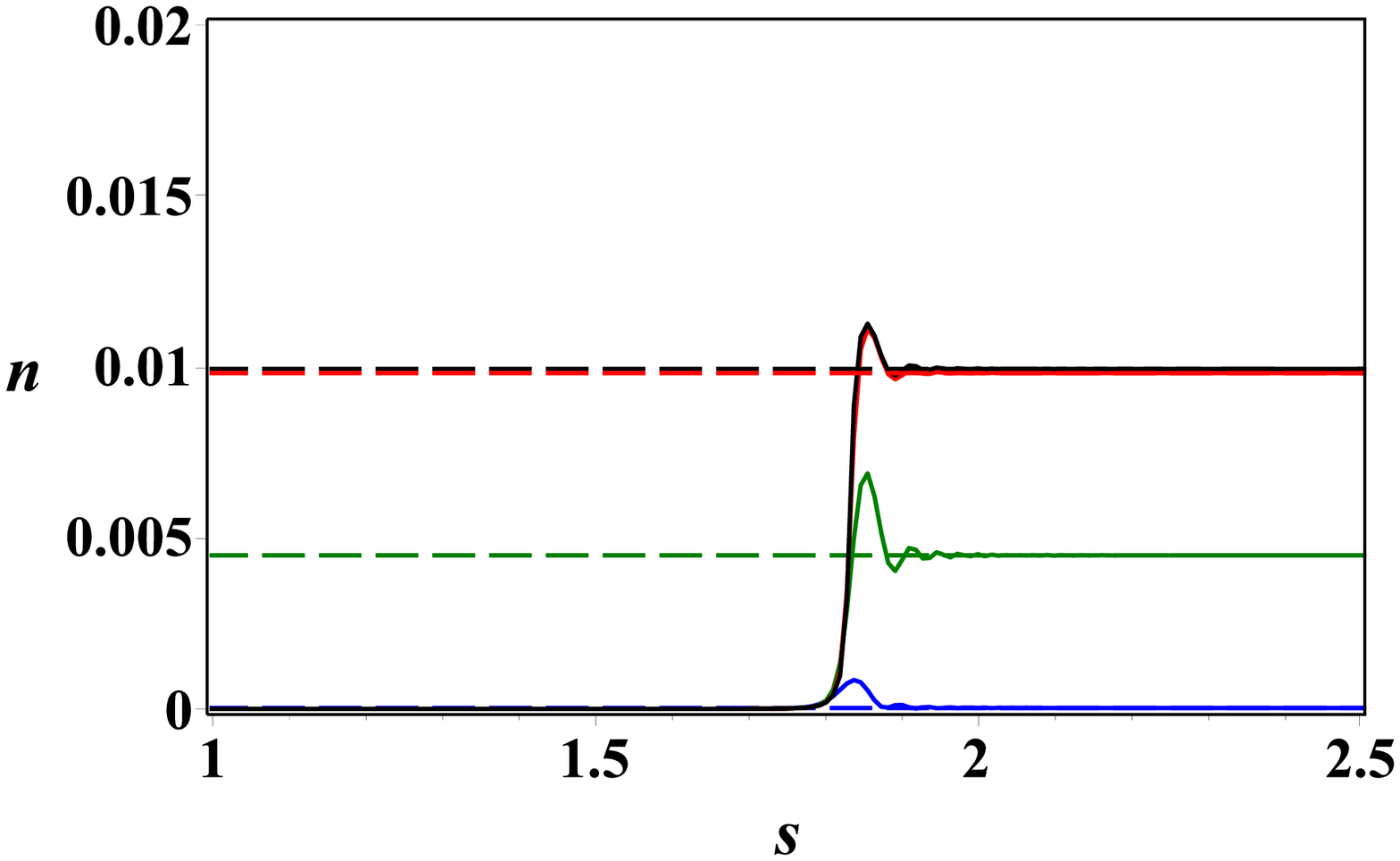}}
(a)
\scalebox{0.325}{\includegraphics{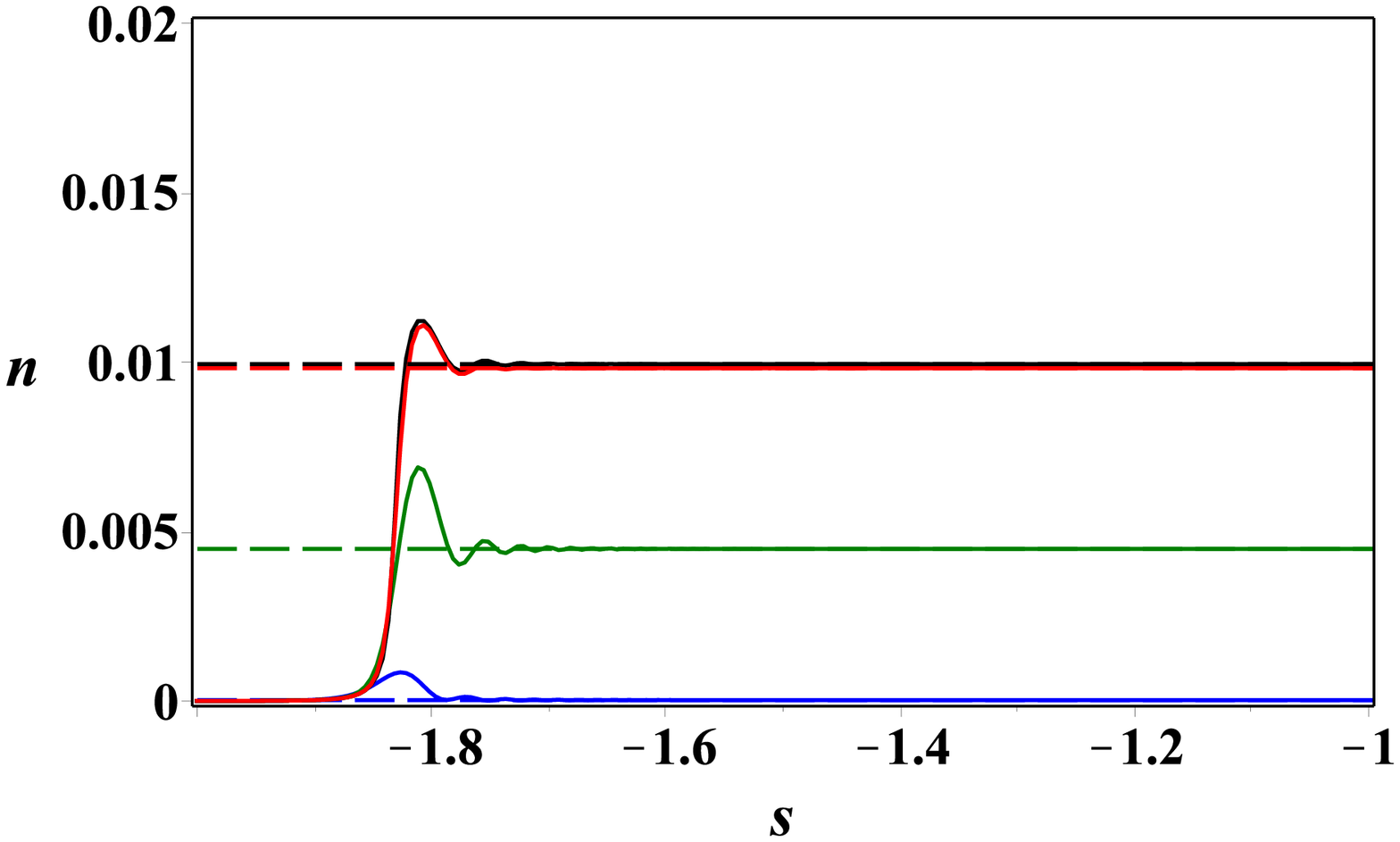}}
(b)
\end{center}
\caption{(Color online) Semi-finite pulse. Density of defects vs. $s$ ($\tau_0 
= 500$, ${h}_0 =20$). Blue line ($N=32$),  green line ($N=64$), red line 
($N=128$), black line ($N=256$). Dashed lines correspond to the 
asymptotic formula (\ref{TL1a}). (a) Paramagnet $\rightarrow$ ferromagnet 
transition. (b) Ferromagnet $\rightarrow$ paramagnet transition. }
\label{DP}
\end{figure}
\begin{figure}[tbp]
\begin{center}
\scalebox{0.325}{\includegraphics{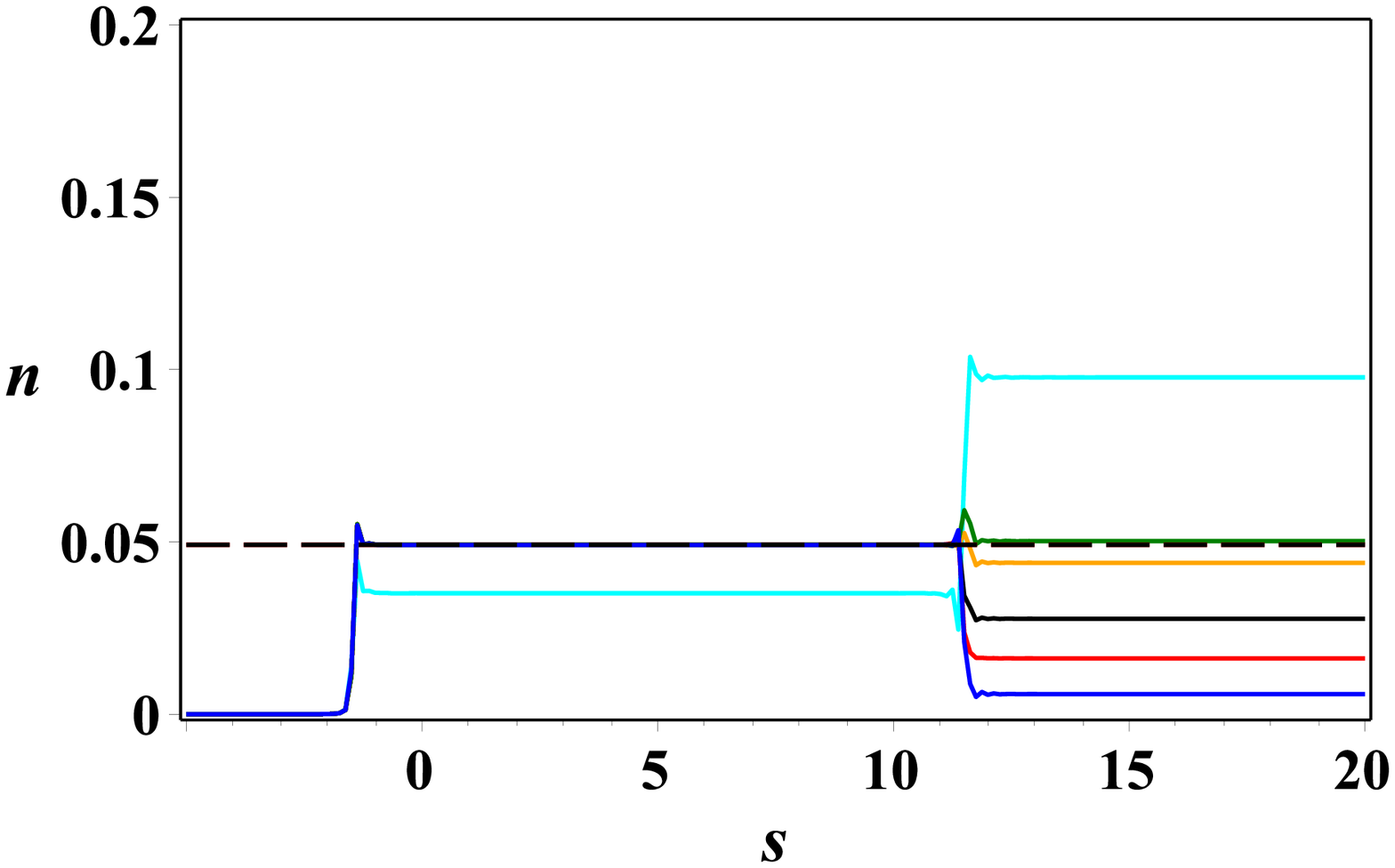}}
(a)
\scalebox{0.325}{\includegraphics{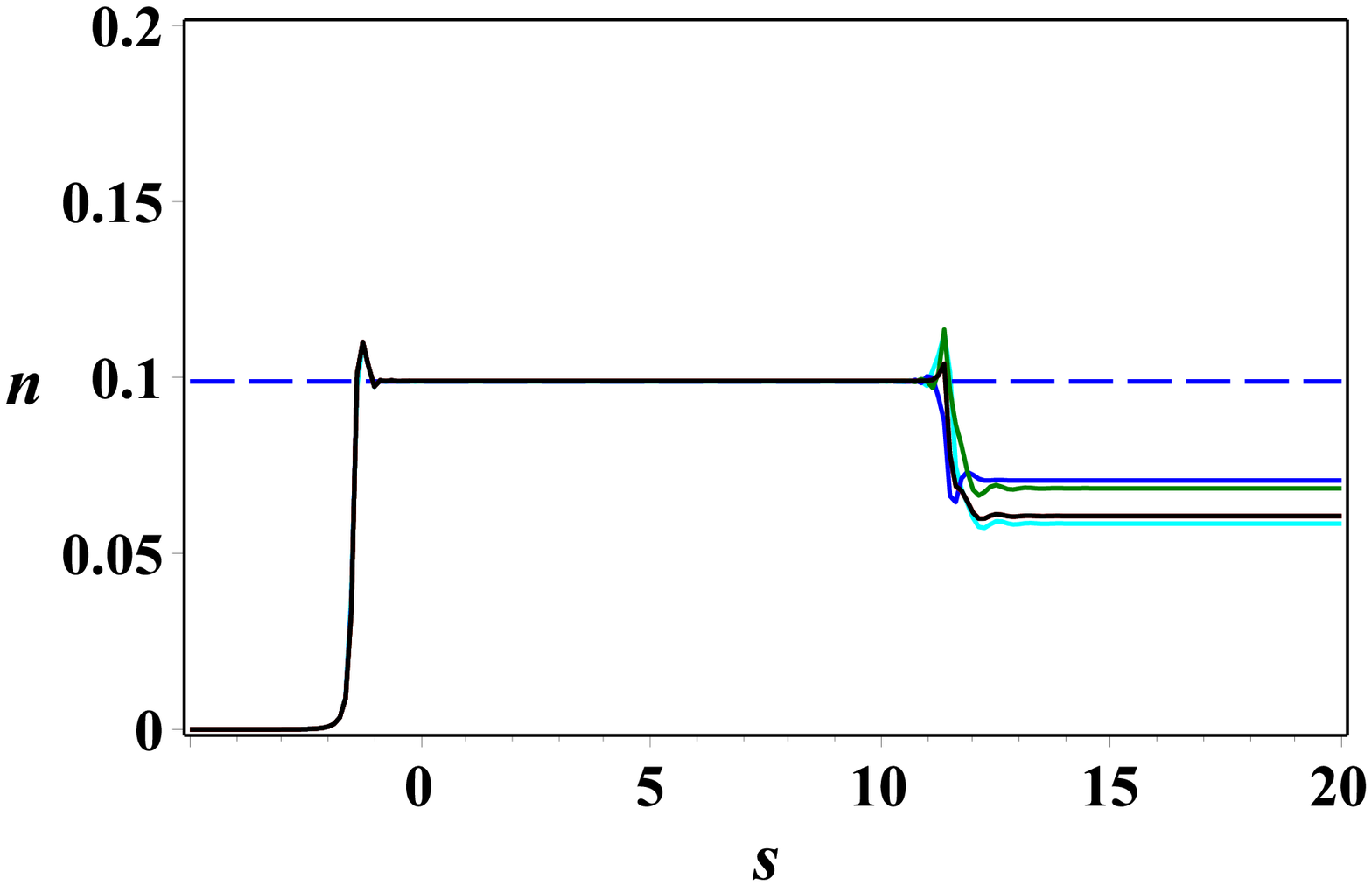}}
(b)
\end{center}
\caption{(Color online) Finite pulse. Density of defects vs. $s$ (${h}_0 =10$, 
$\delta=10$). Cyan line ($N=32$), blue line ($N=48$),  green line ($N=64$), 
red line ($N=128$), black line ($N=256$). Dashed lines correspond to the 
asymptotic formula (\ref{TL1a}). (a) $\tau_0 = 20$. (b) $\tau_0 = 5$. }
\label{DP1}
\end{figure}
\begin{figure}[tbp]
\begin{center}
\scalebox{0.325}{\includegraphics{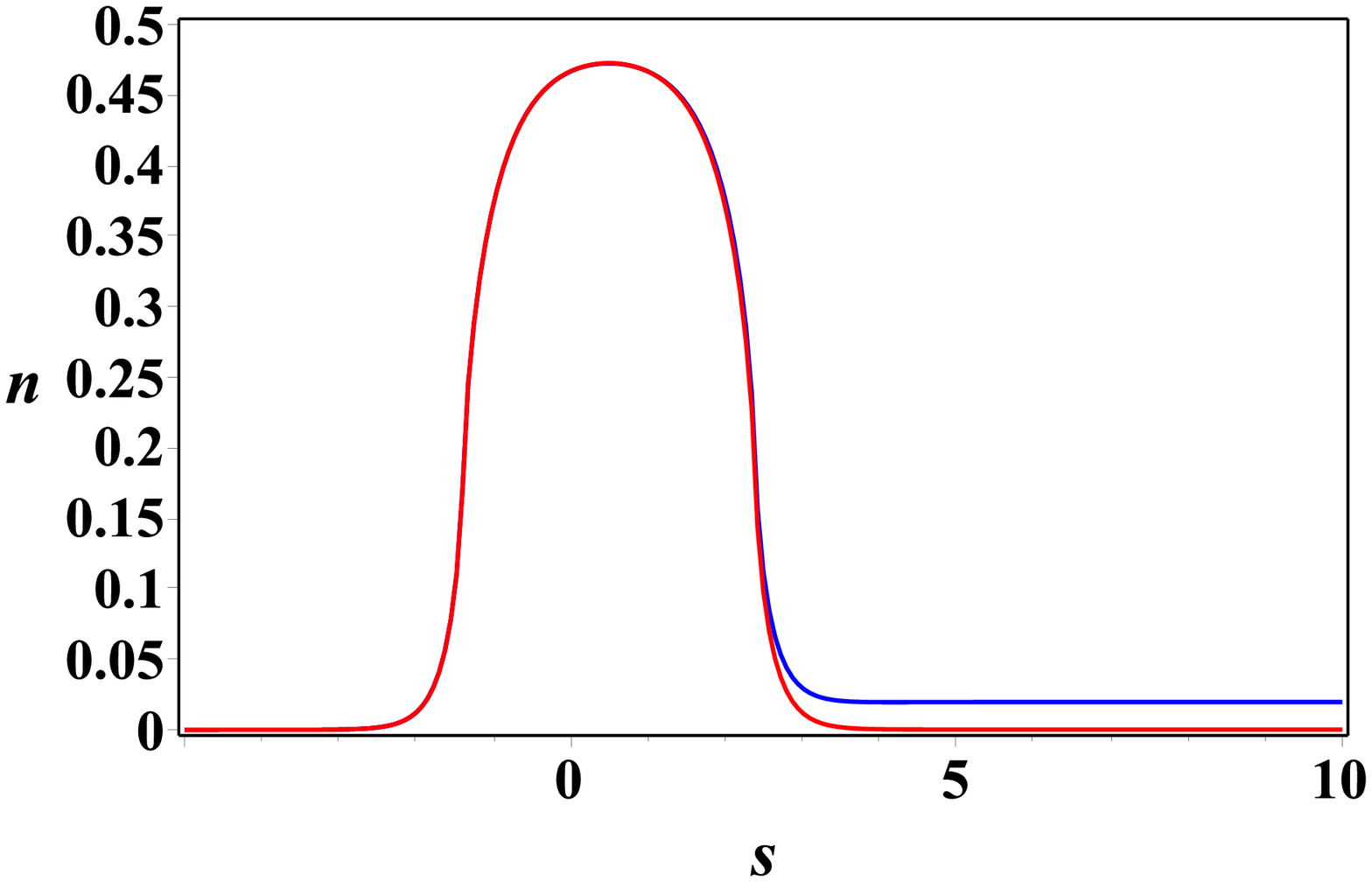}}
(a)
\scalebox{0.325}{\includegraphics{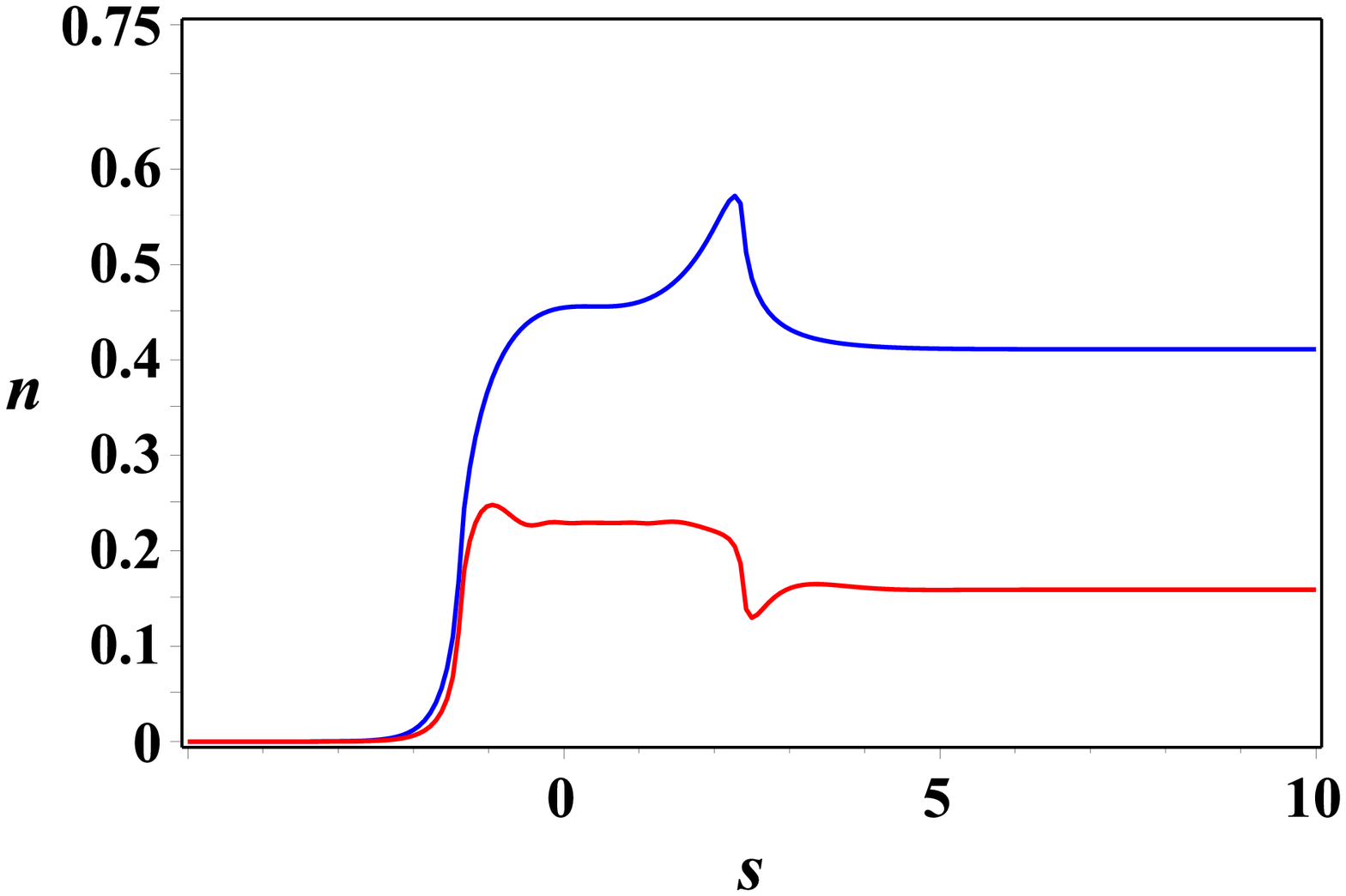}}
(b)
\end{center}
\caption{(Color online) Finite pulse. Density of defects vs. dimensionless 
time $s= t/\tau_0$ for $N=32,564,128,256$ spins (${h}_0 =10$, 
$\delta=1$). (a) $\tau_0 = 0.01$ (blue) , $\tau_0 = 0.001$ (red). (b) $\tau_0 
= 0.1$ (blue) , $\tau_0 = 1$ (red).}
\label{DP2}
\end{figure}

When $N\rightarrow \infty$, the sum in Eq. (\ref{TL1a}) can be replaced by integral, \begin{align}
n= \lim_{N\rightarrow \infty} \frac{\mathcal N}{N}= 1-\frac{1}{2\pi}\int^{\pi}_{-\pi}P_{gs}(\varphi) d\varphi
\end{align}
Substituting  $P_{gs}(\varphi)$ from Eq. (\ref{AEq1}), we find
\begin{align}\label{AEq1c}
	n=  \frac{1}{2\pi}\int^{\pi}_{-\pi}e^{-\pi \nu^2(\varphi)} d\varphi,
\end{align}
where 
\begin{align}
	\nu^2(\varphi) = { J\tau_0}(h_m 
	-1-\sqrt{h_m^2-2h_m\cos\varphi +1}). 
\end{align}
	Next, using the method of 
	steepest descent, we obtain
	\begin{align}
	n= \frac{1}{2\pi\alpha},
	\end{align}
where $\alpha = 2\tau_0 J h_m/|h_m -1|$.

As it was shown in the previous section, during the slow evolution only long 
wavelength modes, with the lowest $\varphi = \pi/N$, can be excited. Thus, 
in the adiabatic regime, $\omega_k^2 \gg 1$, we can approximate the 
average number of defects 
at the end of evolution by LZ formula,
\begin{align}
n=1 -e^{-\pi\omega^2},
\end{align}
where $\omega^2 = \pi^2 \tau_0 J/N^2 $ \cite{DJ,CLDJ}.

In Figs. \ref{DP} -\ref{DP2} we compare  our theoretical predictions with the 
results of  numerical simulations performed for $N=32,48,64,128,256$ 
spins. Solid lines present the results of the numerical simulations and 
dashed lines correspond to the asymptotic formula of Eq. (\ref{TL1a}). We 
assume that initially the system was in the ground state.  One can observe 
that the results predicted by the asymptotic formula (\ref{TL1a}) are in  
good agreement with the numerical results.

\section{Conclusion}

We have studied analytically and numerically the quench dynamics of the quantum Ising chain in a transverse time-dependent magnetic field. We extend the LZ-formula to non-adiabatic evolution of the quantum system. For the adiabatic evolution, our predictions coincides with the results given by the LZ-formula. Numerical simulations show good agreement between analytical and numerical results. 

Under a shock-wave load the dynamics of the system is more complicated, than in the case of the semi-finite pulse. The final state of the system  depends on the amplitude and pulse velocity and may results in a significant density of defects.

\section*{Acknowledgments}

AIN acknowledges the support by the CONACyT.

\end{document}